\documentclass{article}

\usepackage{arxiv}

\usepackage[utf8]{inputenc} 
\usepackage[T1]{fontenc}    
\usepackage{hyperref}       
\usepackage{url}            
\usepackage{booktabs}       
\usepackage{amsfonts}       
\usepackage{nicefrac}       
\usepackage{microtype}      
\usepackage{lipsum}		
\usepackage{graphicx}
\usepackage{doi}
\usepackage{amsmath}
\usepackage{amssymb}
\usepackage{amsthm}
\usepackage{graphicx}

\usepackage{listings}
\usepackage{xcolor}
\usepackage{algorithm}
\usepackage{algorithmic}

\usepackage{enumitem}
\usepackage{graphicx}

\definecolor{codegreen}{rgb}{0,0.6,0}
\definecolor{codegray}{rgb}{0.5,0.5,0.5}
\definecolor{codepurple}{rgb}{0.58,0,0.82}
\definecolor{backcolour}{rgb}{0.95,0.95,0.92}

\lstdefinestyle{mystyle}{
    backgroundcolor=\color{backcolour},   
    commentstyle=\color{codegreen},
    keywordstyle=\color{magenta},
    numberstyle=\tiny\color{codegray},
    stringstyle=\color{codepurple},
    basicstyle=\ttfamily\scriptsize,
    breakatwhitespace=false,         
    breaklines=true,                 
    captionpos=b,                    
    keepspaces=true,                 
    numbers=left,                    
    numbersep=5pt,                  
    showspaces=false,                
    showstringspaces=false,
    showtabs=false,                  
    tabsize=2
}
\title{Quantum Encryption of superposition states with Quantum Permutation Pad in IBM Quantum Computers}

\author{ {Maria Perepechaenko} \\
	Quantropi Inc.\\
	Ottawa, Canada\\
	\texttt{maria.perepechaenko@quantropi.com} \\
	\And
	{Randy Kuang} \\
	Quantropi Inc.\\
	Ottawa, Canada\\
	\texttt{randy.kuang@quantropi.com} \\
}

\renewcommand{\shorttitle}{Encrypting superposition states with QPP algorithm}

\begin{document}
\maketitle

\begin{abstract}
We present an implementation of Kuang and Bettenburg’s Quantum Permutation Pad (QPP) used to encrypt superposition states. The project was conducted on currently available IBM quantum systems using the Qiskit development kit. This work extends previously reported implementation of QPP used to encrypt basis states and demonstrates that application of the QPP scheme is not limited to the encryption of basis states. For this implementation, a pad of 56 2-qubit Permutation matrices was used, providing 256 bits of entropy for the QPP algorithm. An image of a cat was used as the plaintext for this experiment. The plaintext was randomized using a classical XOR function prior to the state preparation procedure. To create corresponding superposition states, we applied a novel operator defined in this paper. These superposition states were then encrypted using QPP, with $2$-qubit Permutation Operators, producing superposition ciphertext states. Due to the lack of a quantum channel, we omitted the transmission and executed the decryption procedure on the same IBM quantum system. If a quantum channel existed, the superposition ciphertext states could be transmitted as qubits, and be directly decrypted on a different quantum system. 
We provide a brief discussion of the security, although the focus of the paper remains on the implementation. Previously we have demonstrated QPP operating in both classical and quantum computers, offering an interesting opportunity to bridge the security gap between classical and quantum systems. This work broadens the applicability of QPP for the encryption of basis states as well as superposition states. We believe that quantum encryption schemes that are not limited to basis states will be integral to a secure quantum internet, to reduce vulnerabilities introduced by using two separate algorithms for secure communication between a quantum and a classical computer. 
\end{abstract}


\section{Introduction}
Recent developments in the field of quantum computing, including demonstrations of quantum supremacy~\cite{quantSupremeOne,quantSupremeTwo}, indicate a significant shift in cryptography. Quantum cryptography and quantum encryption are now critical elements in the field of cryptography. The terms ``Quantum encryption" and "Quantum cryptography" often refer to Quantum Key Distribution (QKD) or Post-Quantum cryptography (PQC). Both of these methods are used to establish a secret key for digital symmetric encryption. However, the former uses quantum mechanics and is implemented in a quantum system with a classical post-processing part, and the latter describes classical algorithms with an underlying mathematical problem that can not be solved by a quantum computing system~\cite{alma991045160358205161,SeitoTakenobu2021TSoP,MaimutDiana2019PCaa}. 


There exists another notable family of algorithms that can be described as Quantum encryption and Quantum cryptography but is not frequently mentioned. This family includes the symmetric cryptographic algorithms to be implemented on quantum systems, such as the quantum implementation of the AES-128. The first quantum circuit design of AES-128 was proposed by Almazrooie \textit{et al.}~\cite{aes-circuit-Almazrooie-2018}. This work was further improved by Langenberg \textit{et al.} to reduce the implementation to 880 qubits, 1507 X gates, 107960 CNOT gates, and 16940 Toffoli gates~\cite{aes-circuit-Langenberg-2020}. Wang, Wei, and Long proposed a different implementation with 656 qubits, 1976 X gates, 101174 CNOT gates, and 18040 Toffoli gates~\cite{aes-circuit-wang-2021}. Zou \textit{et al.} introduced a new implementation with 512 qubits for AES-128~\cite{aes-circuit-zou-2020}. All of these implementations, however, require noticeably large quantum resources not currently available, thus, an efficient implementation of AES-128 might not be possible in the near future. 

Other quantum communication methods proposed in recent years include Quantum secure direct communication without a pre-shared key or QSDC~\cite{QSDCdeng2003,QSDCdeng2004,QSDCzhang2017}, and a lightweight quantum encryption scheme~\cite{quantum-encryption-Hu-2021}. The lightweight quantum encryption scheme uses a generic unitary gate in conjunction with the CNOT gates to enforce diffusion and confusion capability to the cipher quantum states. These cipher quantum states are generally in superposition, thus preventing possible eavesdropping during their transmission to a receiver. Other important algorithms include the Quantum One-Time Pad algorithm (QOTP)~\cite{QOTP,MoscaQOTP,Leung} and the $\mathcal{EHE}$ algorithm~\cite{EHE}. 

A different symmetric cryptographic algorithm was proposed by Kuang and Bettenburg in 2020, called the Quantum Permutation Pad (QPP)~\cite{qpp-kuang2020}. QPP has since been applied to create lightweight block cipher \cite{qpp-aes-kuang-2021}, streaming cipher \cite{qpp-astesj-kuang-2021}, entropy expansion \cite{qpp-expansion-kuang-2021}, and a pseudo-Quantum Random Number Generator or pQRNG \cite{pqrng-kuang-2021}. Kuang and Barbeau recently proposed a concept of universal cryptography using QPP which can be implemented in both classical and quantum computing systems \cite{QPP}. Perepechaenko and Kuang demonstrated the implementation of QPP in the IBM Quantum systems using a few simple examples \cite{EPJ,ICCCAS,JCM}. The QPP uses permutation matrices to encrypt any given plaintext, and their respective Hermitian conjugates to decrypt the produced ciphertext and successfully unveil the plaintext. To guarantee that the decrypting party applies respective conjugates of the same permutation matrices that were used to encrypt plaintext, the two communicating parties pre-share a secret key using some asymmetric key exchange scheme. This secret key is used to create the Permutation Pad, consisting of permutations, as well as the Inverse Permutation Pad, consisting of the Hermitian conjugates of the permutations used for the encryption. Permutation matrices can be efficiently implemented in a quantum circuit using CNOT, NOT, Toffoli, and SWAP gates~\cite{shende2003}. Moreover, permutation matrices can be used to specify Qiskit operator objects using row-by-row matrix description, thus making it possible to implement QPP in the IBM Quantum systems. On the other hand, permutation matrices can be converted into arrays, to be implemented in classical systems. To the best of our knowledge, QPP is the first ever hybrid quantum-classical scheme. It will require further examination to determine whether there are advantages to using the quantum implementation of QPP as opposed to a classical implementation, for instance, we would expect that the quantum implementation is more efficient than the classical implementation. However, one of the main attributes of QPP is that it can be implemented in both classical and quantum devices. Previous implementations of QPP have shown that it is lightweight and can be run today on a free-of-charge IBM Quantum system as well as any classical computer, unlocking the potential of being widely used for quantum encryption in classical systems, quantum systems, as well as hybrid systems, with significant potential use-cases such as quantum internet~\cite{ICCCAS,JCM,EPJ,qpp-aes-kuang-2021,qpp-astesj-kuang-2021,qpp-expansion-kuang-2021}. Having a hybrid encryption scheme reduces the limitations and vulnerabilities of a scenario where two different schemes are used on quantum and classical devices, and therefore we continue advancing QPP to ensure that both quantum and classical QPP implementations are efficient and secure. 

In this work, we present a fully-functioning implementation of the QPP algorithm used to encrypt superposition states with Qiskit in IBM Quantum systems. We were curious to examine whether QPP is limited to basis states, and also explore whether there are any security benefits brought about by introducing superposition states. This work goes beyond academic interest and exhibits potential to be used in the future quantum internet where the states of quantum systems may generally be superposition states transmitted over a quantum channel. This paper builds on and extends our previous research~\cite{EPJ,ICCCAS,JCM}. In~\cite{EPJ,ICCCAS,JCM} we describe how QPP can be used to encrypt basis states in an IBM Quantum device using Qiskit. We find it important to provide the reader with a brief description of the previous work, to facilitate a better understanding of the developments described in this paper. 

\subsection{Encryption of basis states using Quantum Permutation Pad}\label{sec:pureQPP}

QPP is a symmetric encryption scheme that requires a pre-shared secret key for configuration. Consider two communicating parties, Alice and Bob, with a pre-shared secret key. Alice prepares a message "Hello Bob!", expressed as a binary string to be encrypted using QPP with 2-qubit permutations. This QPP scheme uses Permutation Operators that act on $2$-qubit states, one at a time. Thus, Alice splits the message into  $2$-bit blocks, to be encrypted one at a time. For each message block, a quantum circuit is created with $2$ qubits and $2$ classical bits. The initial state of each such circuit is set to reflect the corresponding binary message block using Qiskit's \verb|QuantumCircuit.initialize(Statevector.from_label(state_| \verb|vector))| command. That is, if the corresponding message block is `01', the initial state of the system will be set to $|01\rangle$. Using the pre-shared secret key Alice generates a Permutation Pad consisting of $2$-qubit Permutation Operators, by specifying matrices row by row, and converting them to Operators using Qiskit's \verb|Operator()| command. There are $24$ possible distinct $2$-qubit Permutation Operators, however, to achieve $256$ bits of entropy the permutation pad consists of $56$ $2$-qubit permutation matrices chosen at random using the pre-shared secret key. In this case, repetitions are allowed. To encrypt every plaintext state, that is the state initialized according to the plaintext message block, Alice dispatches a Permutation Operator from the Permutation Pad to act on a given plaintext state, producing a ciphertext state. The dispatching of the Permutation Operator from the Permutation Pad can be done using the pre-shared secret key. 

\begin{figure}[ht]
\includegraphics[width=8 cm]{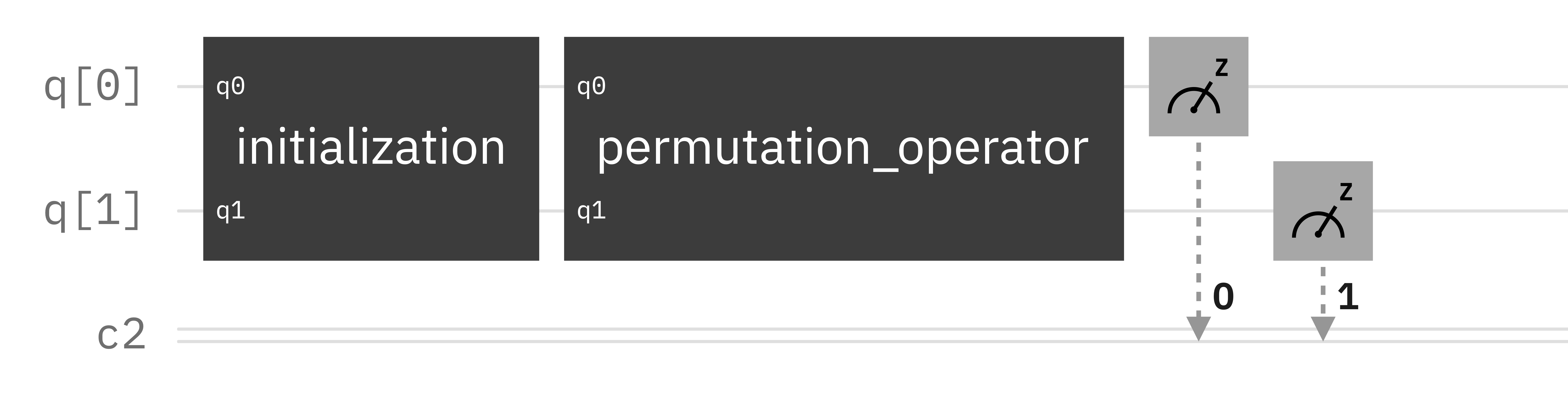}
\centering
\caption{A sample encryption circuit of a plaintext block being encrypted with QPP using Qiskit on an IBM Quantum system. The initialization operator produces an initial state vector that corresponds to a given specified binary string. The Permutation Operator encrypts the state vector to produce a cipher state, which is then measured. This illustration corresponds to our previous work~\cite{ICCCAS,JCM}.}
\label{Fig1}
\end{figure}

Due to a lack of a quantum channel, the quantum ciphertext states can not be sent directly and must be measured, with the highest probability results used to create a binary file containing the ciphertext. Figure~\ref{Fig1}  illustrates a general form of the encryption circuit, used to encrypt each plaintext block. 

\begin{figure}[ht]
\centering
\includegraphics[width=8 cm]{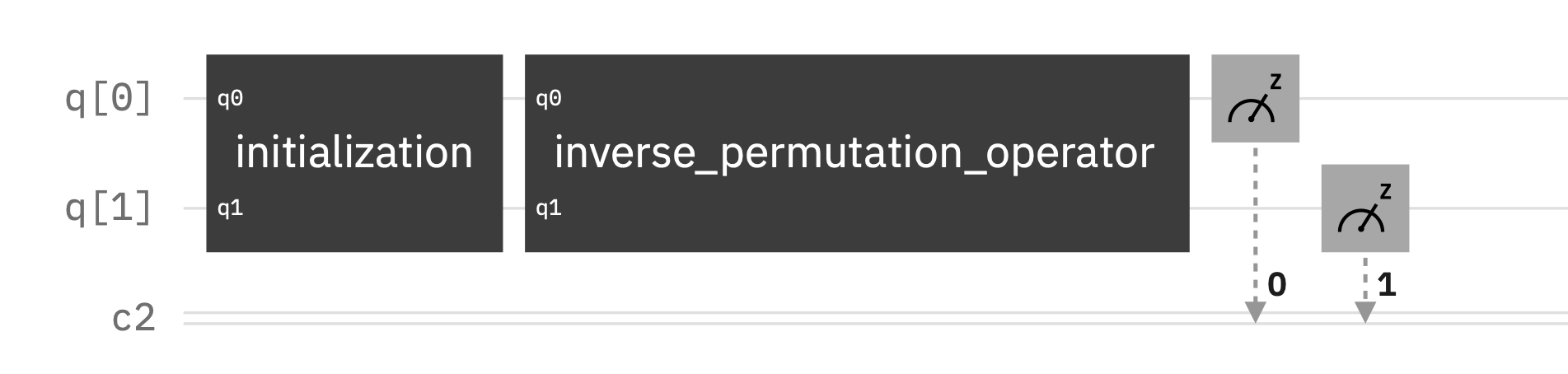} 
\caption{A sample decryption circuit of a ciphertext block being decrypted with QPP using Qiskit on an IBM Quantum system. The initialization operator produces an initial state vector that corresponds to a given specified binary string. The Hermitian conjugate of the Permutation Operator decrypts the state vector to produce a message state, which is then measured. This illustration corresponds to our previous work~\cite{ICCCAS,JCM}.}
\label{Fig2}
\end{figure}

\begin{figure}[ht]
\centering
\includegraphics[width=16cm]{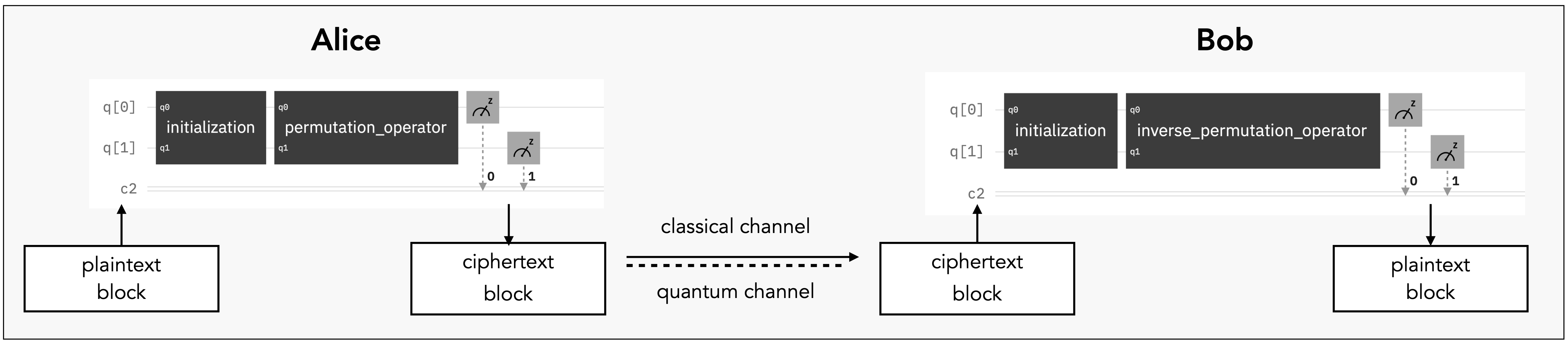}
\caption{A diagram illustrating the logical flow of the QPP encryption and decryption algorithm as described in our previous work~\cite{ICCCAS,JCM}.}
\label{Fig3}
\end{figure}

Upon receiving the binary file containing the ciphertext, Bob extracts the ciphertext as a binary string. Bob then generates an Inverse Permutation Pad, using the pre-shared secret key, which consists of Hermitian conjugates of the corresponding Permutation Operators in the Permutation Pad. Similar to Alice, Bob separates the ciphertext into $2$-bit ciphertext blocks and creates a decryption circuit for each such block. Similar to the encryption circuits, the initial state in each decryption circuit is set to reflect the corresponding ciphertext bits in each given $2$-bit ciphertext block. The ciphertext quantum states are then acted on with the operators from the Inverse Permutation Pad. These Permutation Operators are dispatched using the pre-shared secret key, one for each circuit. The dispatching procedure is the same for Bob and Alice to ensure that Bob dispatches precisely the Hermitian conjugates of the Permutation Operators used by Alice. This process decrypts the ciphertext states and produces plaintext states. The plaintext states are then measured, to yield the binary message. Figure~\ref{Fig2} illustrates the general form of a decryption circuit for each ciphertext block. Figure~\ref{Fig3} depicts the general QPP symmetric scheme. 

\section{Materials and Methods}\label{sec:materialsandmethods}
The motivation for this work was a question of whether the $2$-qubit QPP can be used to encrypt superposition states, rather than basis states, and whether adding the superposition into the framework increases the security. We present an implementation of QPP with 2-qubit Permutation Operators applied to superposition states on the IBM quantum systems using the Qiskit software development kit. We will demonstrate the encryption of the image given in Figure~\ref{Fig4}.

Similar to our previous work~\cite{EPJ,ICCCAS,JCM}, the current implementation is done using 2-qubit Permutations Operators. Such $2$-qubit Permutation Operators act simultaneously on $2$ qubits, in other words, these permutations permute $4$ state vectors. Hence, each 2-qubit permutation is an element of the Symmetric group $S_4$. There are at most $4! = 24$ such 2-qubit permutations. Each such permutation supplies $\log_2 24 \approx 4.58$ bits of entropy. Thus, to achieve $256$ bits of entropy, we need to use $56$ of $2$-qubit permutations chosen at random to encrypt any given plaintext. In this work, we create a Permutation Pad consisting of $56$ of $2$-qubit Permutation Operators used to encrypt a given plaintext. 

Throughout the manuscript, we refer to $2$-qubit QPP as QPP, and $2$-qubit permutations and Permutation Operators as permutations and Permutation Operators respectively, unless explicitly states otherwise. 

We first give a general overview of the procedure and then describe in more detail techniques that are special to the current implementation and were not used in our previous work in Section~\ref{sec:analysisanddiscussion}.

\subsection{Initial Set Up Procedure}
\begin{enumerate}
\item[Step 1]  As discussed in~\cite{QPP}, QPP is a symmetric encryption algorithm that requires a pre-shared secret key. Thus, communication starts with two parties, Alice and Bob, establishing a pre-shared secret. 
\item[Step 2] Alice uses the pre-shared secret to produce $56$ of the $4\times 4$ permutation matrices using the secret key and the Fisher-Yates shuffling algorithm. Once the permutation matrices are created, they can be specified row-by-row to produce quantum operators using Qiskit's \verb|Operator()| command. These Permutation Operators are used to populate a list which we refer to as the Permutation Pad. Similarly, Bob uses the same pre-shared secret to produce an Inverse Permutation Pad, consisting of respective Hermitian conjugates of the Permutation Operators in the Permutation Pad. In order to create the respective Hermitian conjugates of the Permutation Operators, one can simply create the conjugate transpose matrices of each permutation matrix. These conjugate transpose matrices are then specified row-by-row to produce Hermitian conjugate operators using the \verb|Operator()| command.
\item[Step 3] The two parties also use the pre-shared key to dispatch the Permutation Operators from the Permutation Pad and their respective Hermitian conjugates from the Inverse Permutation Pad for encryption and decryption respectively. Note that the dispatching procedure must be agreed upon between Alice and Bob to insure that the Permutation Operators dispatched for decryption are precisely the respective Hermitian conjugates of the Permutation Operators dispatched for encryption for the corresponding states.

For the dispatching procedure described in this work, the pre-shared key was used to create a list of indices, where each index indicates the position of the Permutation Operator in the Permutation Pad and its Hermitian conjugate in the Inverse Permutation Pad respectively. Then, based on the index, a single Permutation Operator is dispatched from the Permutation Pad for encryption of a given state. Similarly, a single Hermitian conjugate operator is dispatched from the Inverse Permutation Pad to decrypt the corresponding ciphertext state. That is, one operator per state.  







\end{enumerate}

\subsection{Encryption Procedure}
The first few steps of the encryption procedure described in this section are similar to those described in Section~\ref{sec:pureQPP}.
\begin{enumerate}
    \item[Step 1] Alice prepares a binary message, which we will refer to as the original plaintext.
    \item[Step 2] In contrast to the process described in Section~\ref{sec:pureQPP}, where the original plaintext is being encrypted directly, here the original plaintext is being randomized first to erase any statistical patterns that can leak the private information.  We refer to the original plaintext that has been randomized as the randomized plaintext. 
    For the current implementation, we utilize classical XOR to randomize the original plaintext. That is, we XOR the original plaintext with the pre-shared secret key and produce the randomized plaintext. Recall, that the original plaintext is written in binary and so is the pre-shared secret key, thus, the randomization procedure can be done using Python command 

\begin{lstlisting}[language = Python]
randomized_message=[str(int(message[i])^int(key_for_xor[i])) for i in range(len(message))]
randomized_message=''.join(randomized_message)
\end{lstlisting}

\noindent where \verb|message| denotes the original plaintext, \verb|key_for_xor| denotes the pre-shared secret key block used for randomization, and \verb|randomized_message| denotes the randomized plaintext. This randomized plaintext is used for encryption. 
\item[Step 3] The randomized plaintext is then split into $2$-bit blocks, to be encrypted one at a time, using Python's \verb|for| loop. The reason for separating plaintext in blocks is the current limitations of quantum devices, such as a number of qubits and Quantum volume.
\item[Step 4] For each such $2$-bit block of randomized plaintext, a quantum circuit is created with $2$ qubits and $2$ classical bits. The state vector of each such circuit is initialized according to the corresponding $2$-bit randomized plaintext block using Qiskit's \verb|QuantumCircuit.initialize(Statevector.from_label(state_vector))| command.
\item[Step 5]  Now that the initial state vector of each circuit reflects the corresponding $2$-bit block of randomized plaintext, they are used to produce superposition states. That is, the initialized state of each system is acted on with the unitary operator \verb|superposition_operator|, which is a Qiskit operator class object created using the matrix
\begin{equation}
\hat{H} = \frac{1}{2}\begin{bmatrix}
    1 & 1& -1 & -1\\
    1 & -1 & -i& i\\
    1 & -1 & i &-i\\
    1 & 1 & 1 & 1
    \end{bmatrix},
\end{equation} that produces a respective superposition state. Qiskit allows to create an operator object by initializing it with a matrix given as a list or a Numpy array. Thus we define an operator \verb|superposition_operator| by specifying the underlying matrix $\hat{H}$ row by row as 

\begin{lstlisting}[language = Python]
superposition_operator = Operator([[1/2,1/2,-1/2,-1/2], [1/2,-1/2,(-1j)/2,(1j)/2], [1/2,-1/2,(1j)/2,(-1j)/2], [1/2,1/2,1/2,1/2]])
\end{lstlisting}

Note that \verb|superposition_operator| is not a secret and can be used by the communicating parties as well as the adversary. 
\item[Step 6] The produced superposition states are then encrypted using Permutation Operators dispatched from the Permutation Pad. Note that a single Permutation Operator is dispatched per circuit from the Permutation Pad to encrypt the corresponding superposition state. This process produces the ciphertext state for each circuit. The ciphertext states are themselves superposition states since Permutation Operators are linear. This is discussed in more detail in Section~\ref{sec:analysisanddiscussion}.
\end{enumerate}

We provide the reader with Figure~\ref{Fig5} illustrating the encryption procedure described in this section. 

\subsection{Transmission} 

Assuming that a quantum channel existed, it would have been possible to send the ciphertext states directly to Bob as qubits, act on them on Bob's side, and then measure the plaintext states at the end. However, currently, there does not exist quantum channel. Measuring the ciphertext states to create a classical binary ciphertext will destroy the superposition states and affect the decryption procedure. Therefore, we pretend as if the quantum channel exists, and continue running the source code to mimic the decryption procedure. Note that the decryption will be performed on the same quantum system as the one used for encryption. 

\subsection{Decryption Procedure}
Suppose that Bob received the ciphertext states one by one and has already assigned a quantum circuit to each of them. Suppose that the initial state vector of each such $2$-qubit and $2$-bit circuit reflects the respective ciphertext state vectors. 
\begin{enumerate}
    \item[Step 1] Bob acts on each initial state with the corresponding Permutation Operator dispatched from the Inverse Permutation Pad. This procedure decrypts the encrypted superposition states and produces the original superposition states that correspond to the randomized plaintext states.
    \item[Step 2] Then, Bob acts on the superposition states with the Hermitian conjugate of the \verb|superposition_operator|, producing randomized plaintext states. The Hermitian conjugate of the \verb|superposition_operator| can be produced by specifying conjugate transpose of the matrix $\hat{H}$ row-by-row into Qiskit's \verb|Operator()| command.
    \item[Step 3] These randomized plaintext states are then measured, producing randomized plaintext bits.
    \item[Step 4] The randomized plaintext bits, then, undergo a classical de-randomization procedure using the classical XOR function and the secret key. This process produces the original plaintext. 
\end{enumerate}

We provide the reader with Figure~\ref{Fig6}, illustrating the decryption procedure described in this section. 

\subsection{ENT Randomness testing}\label{sec:ENTtest}
The security of the QPP symmetric scheme is studied thoroughly in Kuang and Barbeau's work~\cite{QPP}. Here, we will use the randomness testing tool ENT to examine if the ciphertext demonstrates good randomness. The ENT program applies various tests to sequences of bytes stored in files and reports the six results of those tests. The program is useful for evaluating pseudo-random number generators for encryption and statistical sampling applications, compression algorithms, and other applications where the information density of a file is of interest. Note that there are various test programs to verify the randomness of any given data, including the Diehard, the NIST SP 800-22, and their combination the Dieharder test. However, only the ENT testing tool works with smaller binary files being a few KB. To perform randomness testing with NIST SP 800-22 the required file size should be at least 100 MB. Given the current free of charge IBMQ, generating files of this size requires a very long running time. In this experiment, the original plaintext is merely 9 KB and so is the ciphertext. Thus, we use the ENT to examine the randomness of plaintext, the randomized plaintext, and the superposed randomized plaintext and ciphertext. Note that the most sensitive testing report of ENT is the Chi Square value, which can detect very small biases at the byte and bit level. Looking at the value of the ENT test, including the Chi Square values, we can see the randomness change at different steps of the QPP encryption.

\section{Results}\label{sec:results}
In this section we provide implementation results of the procedure described in Section~\ref{sec:materialsandmethods}.
We present the results at different stages of the implementation to dynamically show the plaintext and the ciphertext changes. 

Suppose that Alice wants to send an image illustrated in the Figure~\ref{Fig4} to Bob. The encrypting procedure is as described in Section~\ref{sec:materialsandmethods}.  The source code of the encryption procedure is given in the Appendix.

\subsection{The ciphertext}\label{sec:ciphertext}

We want to confirm that the ciphertext is truly random and does not leak any information about the original plaintext. We also want to see dynamic results of how the original plaintext changes at every stage of the ciphertext creation procedure. For that, we provide the reader with Table~\ref{ent} reporting on the ENT randomness test results of the system at different stages of the encryption procedure, namely, the original plaintext, the randomized plaintext, the superposition state, and the encrypted superposition state or the ciphertext. 

To run the ENT tests on the original plaintext and the randomized plaintext, they were converted into binary files. To test the superposition states, the states were measured and the per-shot measurement bitstrings were saved using Qiskit's \verb|job = execute(my_circ,|
\verb|backend=qcomp, shots = 20000, memory = True)| command. This binary information was used for the ENT tests. Notice that measurement destroys the superposition states. Thus, in order to test the ciphertext, we re-run the source code and measure the ciphertext states. The ciphertext states are themselves superposition states, thus, here too, we save the per-shot measurement results and use them for the ENT testing. 

Table~\ref{ent} shows that the sensitive Chi-square, denoted $\chi^2$, reports are 2403.83 for original plaintext indicating very biased and 283.07 for randomized plaintext demonstrating less biased; 254.31 after superposition operations and 262.26 for ciphertext showing good randomness with $\chi^2$ values very close to the optimal 256, respectively. Another interesting report needed to mention is the Serial Correlation Coefficient. Of all reports from the original plaintext to the ciphertext, the ciphertext encrypted with QPP demonstrates the least correlation for each byte to its previous byte. Overall, QPP encryption offers good randomness in its ciphertext. 

In Figure~\ref{Fig7}, we include an image of a ciphertext associated with the plaintext illustrated in Figure~\ref{Fig4}. 

\subsection{Applying Hermitian conjugate of the superposition operator to the ciphertext state}\label{sec:attack}
In the current implementation, the superposition operator $\hat{H}$ is publicly known. Thus, one might argue that the adversary can use it to try and perpetrate an attack to obtain information about the plaintext or the secret key. However, we claim that this scenario does not benefit the attacker without any knowledge of the permutation operator $P$ used for encryption. First, for this attack, we must assume that the adversary has a way to apply the Hermitian conjugate of $\hat{H}$ to the ciphertext without altering it. Note that this is a rather ambitious assumption to make since that would require the adversary to act on the entire ciphertext in its original superposition form, rather than the measurement result. In general, the adversary can measure the ciphertext and obtain a single measurement result corresponding to one of the states $|00\rangle, |01\rangle, |10\rangle,$ or $|11\rangle$. Moreover, since the adversary does not know which permutation operator was used for encryption, they might even further obscure the plaintext by applying the operator $\hat{H}$ to the ciphertext.

In this section, we use ENT tests to verify that after applying the Hermitian conjugate of $\hat{H}$ to the ciphertext, it remains random. In the later sections, we consider this scenario in more detail and discuss whether there are any vulnerabilities introduced by the publicly known $\hat{H}$.

In Section~\ref{sec:createsuperposition}, we discuss that there are only $4$ possible superposition states that correspond to the superposition operator $\hat{H}$ applied to a given input state vector. After a superposition state was acted on with a Permutation Operator it might not be of the same form as the mentioned $4$ possible superposition states up to a global phase. That is if $P$ denotes a Permutation Operator, and $\hat{H}$ the superposition operator, then it is often the case that the ciphertext 
\begin{equation}
    P(\hat{H}|r\rangle) \ne \hat{H}|s\rangle,
\end{equation}up to a global phase, where $r$ and $s$ are in $\{00, 01, 10, 11\}$. Therefore, applying the Hermitian conjugate of $\hat{H}$ to the ciphertext state might not produce a basis state, but rather a different superposition state. Indeed, it might be often the case that 
\begin{equation}
    \hat{H}^{\dagger}(P(\hat{H}|r\rangle))) \ne \hat{H}^{\dagger}(\hat{H}|s\rangle) = |s\rangle,
\end{equation} up to a global phase.
In this case, adversary measuring the state $\hat{H}^{\dagger}(P(\hat{H}|r\rangle)))$ will not gain any advantage. Moreover, it is important to mention that the attacker does not have any knowledge of the permutation $P$ used during encryption. That is, the attacker can measure the result of the operation $\hat{H}^{\dagger}[P(\hat{H}|r\rangle)],$ where $P$ is such that $P(\hat{H}|r\rangle) = e^{i\theta}\hat{H}|r\rangle$, while the same result can be obtained by measuring $\hat{H}^{\dagger}[P(\hat{H}|s\rangle)] = \frac{1}{2}|r\rangle + \frac{1}{2}|s\rangle$ for basis states $r,s \in \{00,01,10,11\}.$ 

Suppose that the adversary knows the superposition operator $\hat{H}$ and  uses its Hermitian conjugate operator $\hat{H}^\dagger$ to act on the superposition state. We are wondering, whether the adversary will gain some information about the plaintext or the secret key. We perform the measurements after $\hat{H}^\dagger$ operation and then test the randomness of the measurement results with ENT. Table~\ref{entattack} illustrates the ENT test results of the mentioned superposition states which were acted on with the operator $\hat{H}^{\dagger}$ as well as the ciphertext, the randomized plaintext, and the optimal values. It can be seen that the reports from the 2nd last column still demonstrate good randomness, especially its $\chi^2$ value. That means, QPP encryption still offers good protection on the original plaintext. 

Table~\ref{entattack} shows that the ciphertext remains random even if the adversary acts on it with the Hermitian conjugate of the operator $\hat{H}$. Note also, that the scenario described in this section is in favor of the adversary. 

\subsection{Result of the encryption and decryption using Quantum Permutation Pad}\label{sec:full}
\begin{figure}[ht]
\centering
\includegraphics[width=5cm]{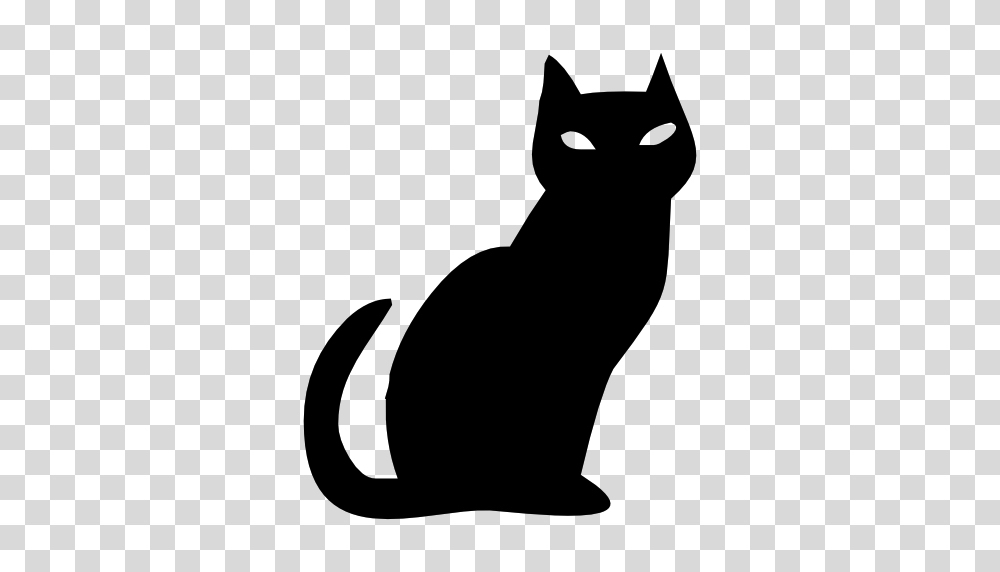}
\caption{An image of the cat to be encrypted with Quantum Permutation Pad as described in this work.}
\label{Fig4}
\end{figure}

\begin{figure}[ht]
\centering
\includegraphics[width=16cm]{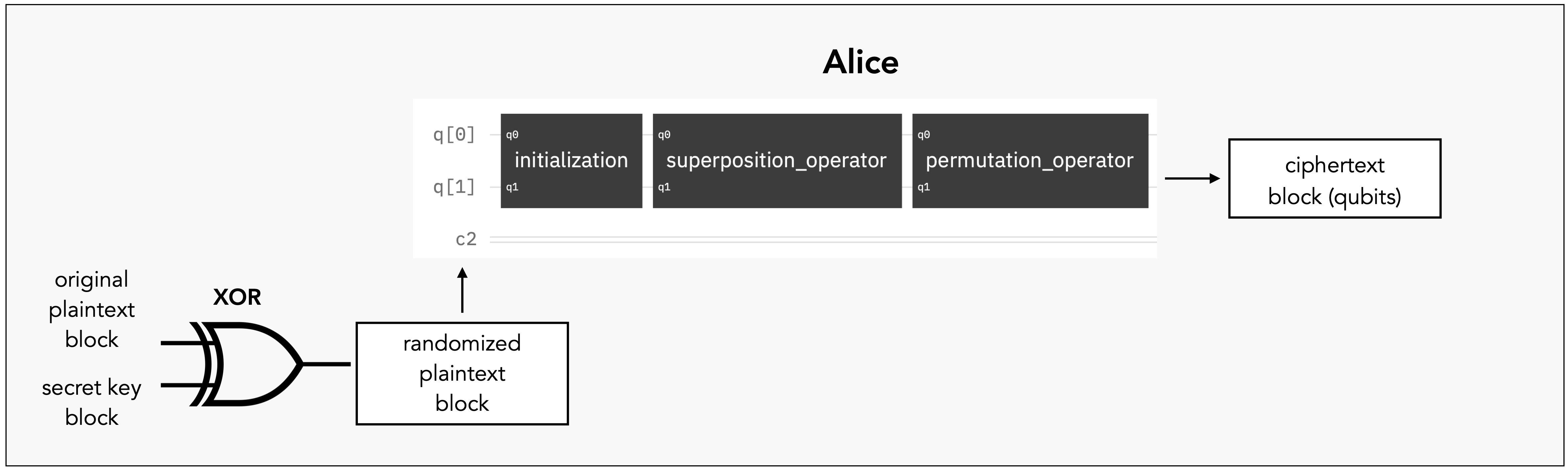}
\caption{A diagram illustrating the logical flow of the encryption procedure of a single plaintext block with QPP using Qiskit on a IBM Quantum system. This illustration corresponds to our current work described in this paper.}
\label{Fig5}
\end{figure}

\begin{figure}[ht]
\centering
\includegraphics[width=16cm]{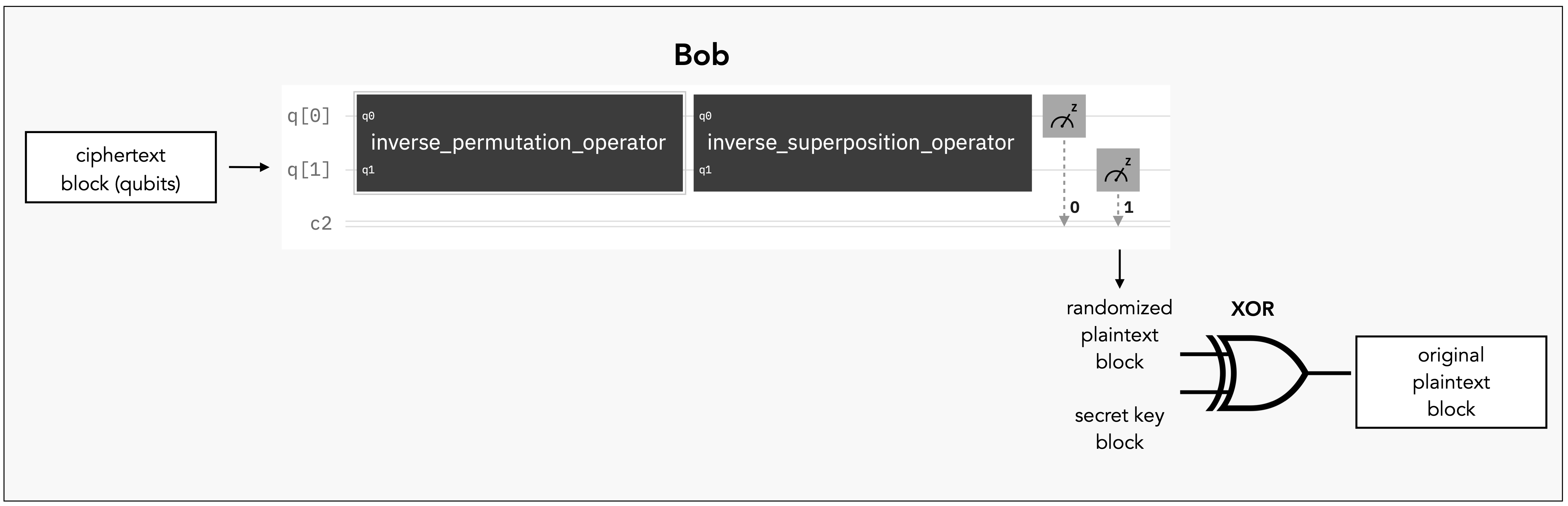}
\caption{A diagram illustrating the logical flow of the decryption procedure of a single ciphertext state with QPP using Qiskit on a IBM Quantum system. This illustration corresponds to our current work described in this paper.}
\label{Fig6}
\end{figure}

\begin{figure}[ht]
\centering
\includegraphics[width=3cm]{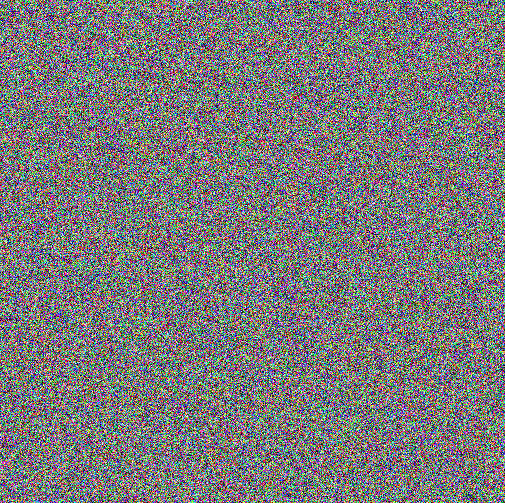}
\caption{An illustration of the ciphertext associated with the plaintext picture in Figure~\ref{Fig4} and the encryption procedure described in Section~\ref{sec:materialsandmethods}.}
\label{Fig7}
\end{figure}

We have run the source code given in Appendix to encrypt and decrypt the image given in Figure~\ref{Fig4}. We invite the reader to test the code on any free-of-charge IBM Quantum system with the same or a different image. The produced ciphertext corresponds to the one described in Section~\ref{sec:attack}. The decryption procedure returns the exact image as in the Figure~\ref{Fig4}.

\begin{table}[h!] 
\centering
\caption{\label{ent}This table illustrates ENT randomness test results of the original plaintext, the randomized plaintext, the superposed randomized plaintext, and the ciphertext against the optimal parameters' values.}
\scalebox{0.72}{
\begin{tabular}{cccccc}
\hline
\textbf{Parameters}	& \textbf{Optimal values}	& \textbf{Original plaintext} & \textbf{Randomized plaintext} & \textbf{Superposition states} & \textbf{Ciphertext}\\
\hline
Entropy		& 7.999999			& 7.864182 & 7.981399 & 7.983191 & 7.982688\\
Chi-square		& 256			& 2403.83 & 283.07 & 254.31 & 262.26\\
Arithmetic Mean & 127.5 & 122.9132 & 127.2537 & 126.4969& 127.7734\\
Monte-Carlo $\pi$ & 3.141592653& 3.213507625 & 3.089324619 & 3.102396514 & 3.135076253\\
Serial Correlation Coefficient &  0.0 & 0.088383 & -0.024435 &  0.001690& 0.000079\\
\hline
\end{tabular}}
\end{table}

\begin{table}[h!] 
\centering
\caption{\label{entattack}This table illustrates ENT randomness test results of the randomized plaintext, the ciphertext, and the ciphertext that was acted on with Hermitian conjugate of the superposition operator against the optimal parameters' values.}
\scalebox{0.72}{
\begin{tabular}{cccccc}
\hline
\textbf{Parameters}	& \textbf{Optimal values} & \textbf{Randomized plaintext} & \textbf{The ciphertext that was acted on with $\hat{H}^{\dagger}$} & \textbf{Ciphertext}\\
\hline
Entropy		& 7.999999			 & 7.981399 & 7.983467 & 7.982688\\
Chi-square		& 256			& 283.07 & 250.18 & 262.26\\
Arithmetic Mean & 127.5 & 127.2537 & 126.5289& 127.7734\\
Monte-Carlo  $\pi$ & 3.141592653& 3.089324619 & 3.130718954 & 3.135076253\\
Serial Correlation Coefficient &  0.0 & -0.024435 &  -0.025627& 0.000079\\
\hline
\end{tabular}}
\end{table}

\section{Analysis and Discussion}\label{sec:analysisanddiscussion}

In this work, we explore whether superposition states can be encrypted using the QPP algorithm with Qiskit on the IBM Quantum systems, and what are the implications of introducing the superposition in the framework of the QPP algorithm. Our previous work~\cite{ICCCAS,JCM,EPJ} described encryption of basis states with QPP, which can be considered a quantum counterpart of the classical QPP encryption algorithm from~\cite{QPP}. We were then curious to examine whether the quantum implementation of QPP can be extended to the framework that can not be replicated on a classical computer, such as encryption of the superposition states and entangled states using QPP. We were also wondering if implementing the QPP algorithm to act on superposition states or entangled states improves security. For one, we were curious whether the inherent randomness nature of superposition adds a new layer of security to the scheme. For this work, we focused only on the superposition states. 

Recall, that in the framework of the QPP symmetric scheme, the plaintext is segmented into $2$-bit blocks due to the limitations of the current quantum computers. Each block is encrypted and decrypted, one at a time, using Python's \verb|for| loop. Thus, at any given moment the quantum system has only one quantum circuit to execute. Each such circuit has $2$ qubits and $2$ bits, which implies that the system can be in one of the $4$ possible state vectors. Therefore, it is sufficient to discuss QPP in the context of a single sample circuit that is in one of the $4$ possible states. 

In this section, we focus mainly on the techniques that are special to the current implementation and were not used in our previous work.

\subsection{Initial Set Up Procedure}

Recall, that communicating parties must first pre-share a secret key $k$ to effectively communicate using the QPP algorithm. As with most other symmetric schemes, the security of QPP does not depend entirely on the design of the scheme itself, but also the secure establishment of the pre-shared secret $k$ as well as the randomness of $k$. That is, we require that $k$ is truly random. To generate $k$ a good source of QRNG can be used, and the randomness of $k$ can be tested using the ENT randomness test. Moreover, we require that $k$ is established using a quantum-safe algorithm. NIST has recently announced candidate algorithms for quantum-safe key encapsulation and digital signatures~\cite{NISTStandards}. These algorithms can be used for authentication and key establishment. Moreover, novel PQC algorithms such as MPPK/DS and MPPK can be used to pre-share a secret key $k$~\cite{MPPK,MPPKDS}.

\subsection{Randomization}\label{sec:ADrandomization}
In the framework of QPP, the plaintext must be randomized prior to being encrypted to eliminate any statistical patterns that can be used by the adversary for statistical analysis attacks. The importance of this step, from the security perspective, is discussed in~\cite{QPP}. However, in our previous work~\cite{ICCCAS,JCM} we did not incorporate a randomization step as we were exploring whether QPP can be implemented in quantum computers and did merely a toy example of the implementation.  

For this implementation, we use the classical XOR function and the secret key to randomize the plaintext. This step is done classically before any quantum circuits are created, as discussed in Section~\ref{sec:materialsandmethods}. However, ideally, the randomization step would be done using the quantum Controlled Pauli X (CX) gates. The CX gates are a quantum alternative to the classical XOR function and would be an ideal tool for the quantum randomization procedure. In this case, the original plaintext as well as the secret key would be broken into $2$-bit blocks, used to initialize the initial states of the corresponding circuits. The circuits in this case would have $4$ qubits and $2$ classical bits, where $2$ qubits correspond to the plaintext, and the other $2$ qubits correspond to the secret key. The secret key qubits would then be used as control qubits, and the X gates would act on the plaintext qubits.

Current limitations, however, make it difficult to implement the above-discussed quantum randomization procedure. We use free-of-charge IBM Quantum computers for this implementation that has a Quantum Volume (QV) of 32. This means that the width (the number of qubits) and the depth (the number of layers) of the largest quantum volume circuits that can be executed ‘successfully’ on a specified quantum device is $5$. We noticed that running QPP circuits with $4$ qubits on devices with QV of $32$ produces a lot of noise and makes it difficult to interpret the correct results. 


\subsection{Creation of superposition states}\label{sec:createsuperposition}
For a single qubit or equivalently $2$ states, it is customary to use the Hadamard gate to create the superposition of $2$ states. Indeed, 
\begin{equation*}
    H|0\rangle = \frac{1}{\sqrt{2}}|0\rangle + \frac{1}{\sqrt{2}}|1\rangle \text{ and } H|1\rangle = \frac{1}{\sqrt{2}}|0\rangle - \frac{1}{\sqrt{2}}|1\rangle,
\end{equation*}
where the amplitudes satisfy $|\alpha|^2 + |\beta|^2 = 1.$ 

\begin{figure}[ht]
\centering
\includegraphics[width=3cm]{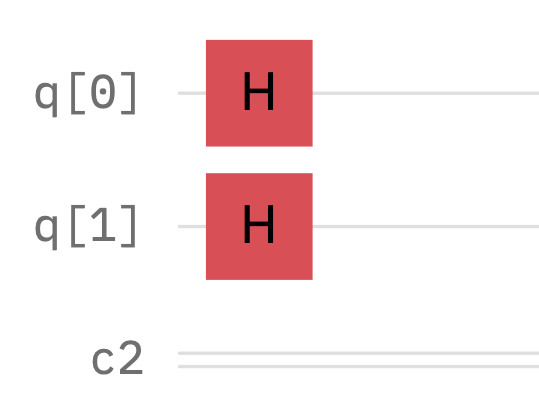}
\caption{A circuit illustrating how to create superposition of four states using Hadamard gates.}
\label{Fig8}
\end{figure}

\begin{figure}[ht]
\centering
\includegraphics[width=10cm]{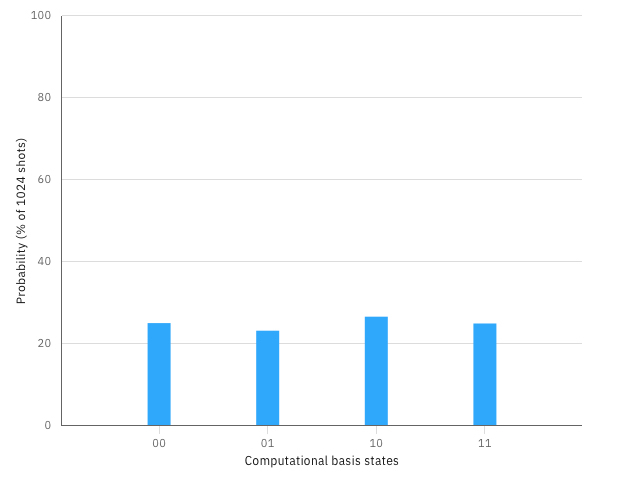}
\caption{A plot histogram of counts result from a circuit execution of the operator $H \otimes H$ applied to the state vector $|00\rangle$.}
\label{Fig9}
\end{figure}

However, for $2$ qubits or $4$ states, there are various ways to create the superposition state. For instance, one could apply Hadamard gate to each qubit, as illustrated in Figure~\ref{Fig8}. In this case the operator acting on the entire system is  
\begin{equation}
    H \otimes H = \frac{1}{2}\begin{bmatrix}
    1 & 1& 1 & 1\\
    1 & -1 & 1& -1\\
    1 & 1 & -1 &-1\\
    1 & -1 & -1 & 1
    \end{bmatrix}.
\end{equation}
The reader can easily verify that this operator creates a superposition for any input state. In Figure~\ref{Fig9} we provide a sample plot histogram of counts result from a circuit execution of the operator $H \otimes H$ applied to the state vector $|00\rangle$. The circuit was executed $1024$ times.


Another way to create a superposition of four states is to act on the state vector of the system with the operator which diagonalizes a permutation operator. In this paper, we use the following superposition operator
\begin{equation}
\hat{H} = \frac{1}{2}\begin{bmatrix}
    1 & 1& -1 & -1\\
    1 & -1 & -i& i\\
    1 & -1 & i &-i\\
    1 & 1 & 1 & 1
    \end{bmatrix}.
\end{equation}
established from the diagonalization of the permutation operator 
\begin{equation*}
    P_1 =
\begin{bmatrix}
    0 & 1 & 0 & 0 \\
    0 & 0 & 0 & 1 \\
    1 & 0 & 0 & 0 \\
    0 & 0 & 1 & 0
\end{bmatrix}.
\end{equation*}
It is easy to prove that the operator $\hat{H}$ is a unitary, and so we leave the proof to the reader.  
Let the state $|rs\rangle$ denote the tensor product $|r\rangle \otimes |s\rangle$, for some $r,s \in \{0,1\}.$ Operator $\hat{H}$ produces the following superposition states depending on the input state 
\begin{equation}
    \hat{H}|00\rangle 
    = \frac{1}{2}|00\rangle + \frac{1}{2}|01\rangle +\frac{1}{2}|10\rangle+\frac{1}{2}|11\rangle
\end{equation}
\begin{equation}
    \hat{H}|01\rangle = 
    \frac{1}{2}|00\rangle - \frac{1}{2}|01\rangle -\frac{1}{2}|10\rangle+\frac{1}{2}|11\rangle
\end{equation}
\begin{equation}
    \hat{H}|10\rangle 
    = \frac{1}{2}|00\rangle - \frac{1}{2}i|01\rangle +\frac{1}{2}i|10\rangle+\frac{1}{2}|11\rangle
\end{equation}
\begin{equation}
    \hat{H}|11\rangle 
    = -\frac{1}{2}|00\rangle + \frac{1}{2}i|01\rangle -\frac{1}{2}i|10\rangle+\frac{1}{2}|11\rangle,
\end{equation}
where the amplitudes satisfy $\sum_{i}|\alpha_i|^{2} = 1.$

\begin{figure}[ht]
\centering
\includegraphics[width=12 cm]{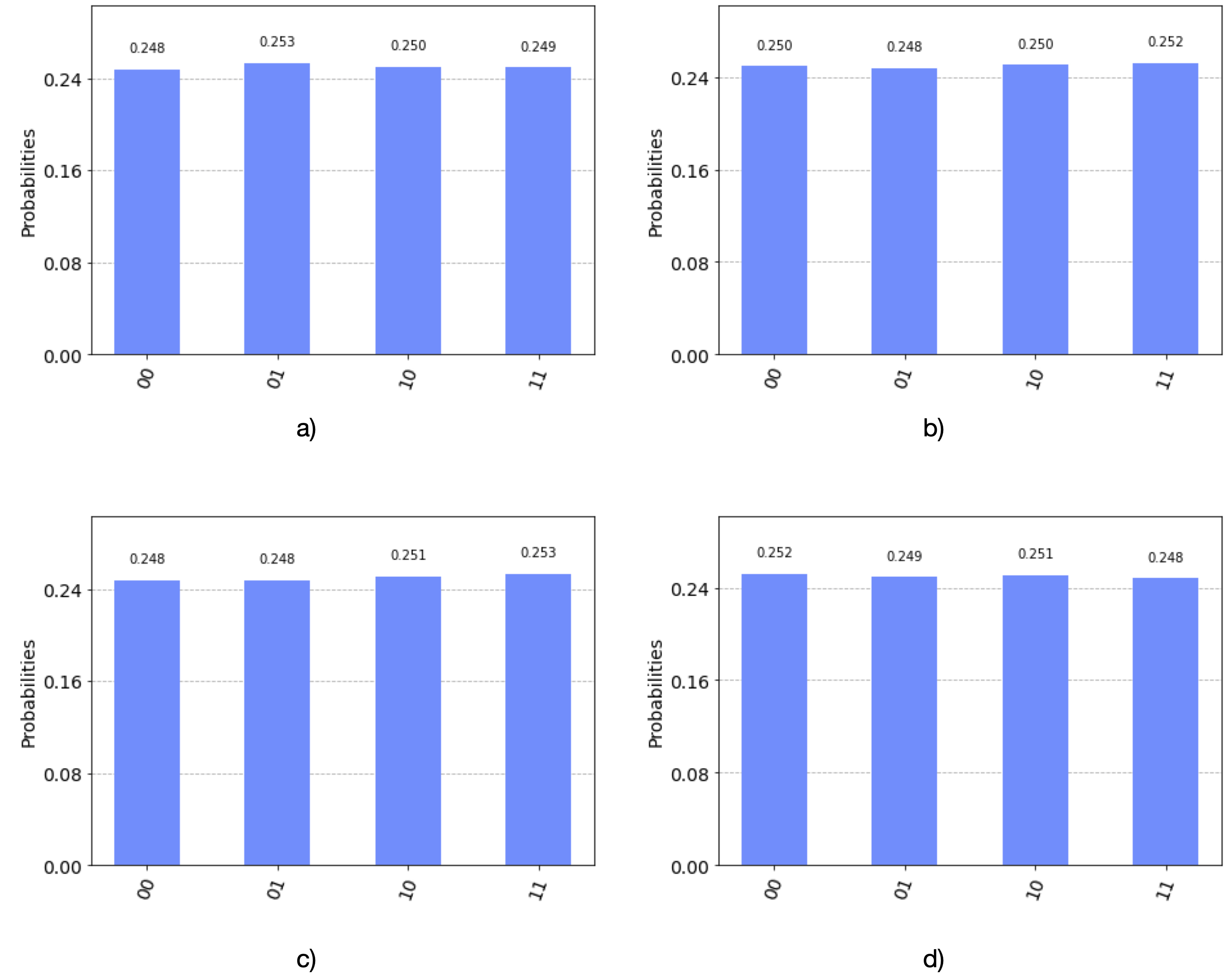}
\caption{This figure illustrates the measurement results in the form of plot histogram of the state $\hat{H}|r\rangle$ for every $r \in \{00,01,10,11\}$. Each corresponding circuit was executed 20,000 times on the IBM Qasm simulator. (\textbf{a}) Plot histogram of measurement results of the state $|\phi\rangle = \hat{H}|00\rangle.$ (\textbf{b}) Plot histogram of measurement results of the state $|\phi\rangle = \hat{H}|01\rangle.$ (\textbf{d}) (\textbf{c}) Plot histogram of measurement results of the state $|\phi\rangle = \hat{H}|10\rangle.$ (\textbf{d}) Plot histogram of measurement results of the state $|\phi\rangle = \hat{H}|11\rangle.$}
\label{Fig10}
\end{figure}

\begin{figure}[ht]
\centering
\includegraphics[width=12 cm]{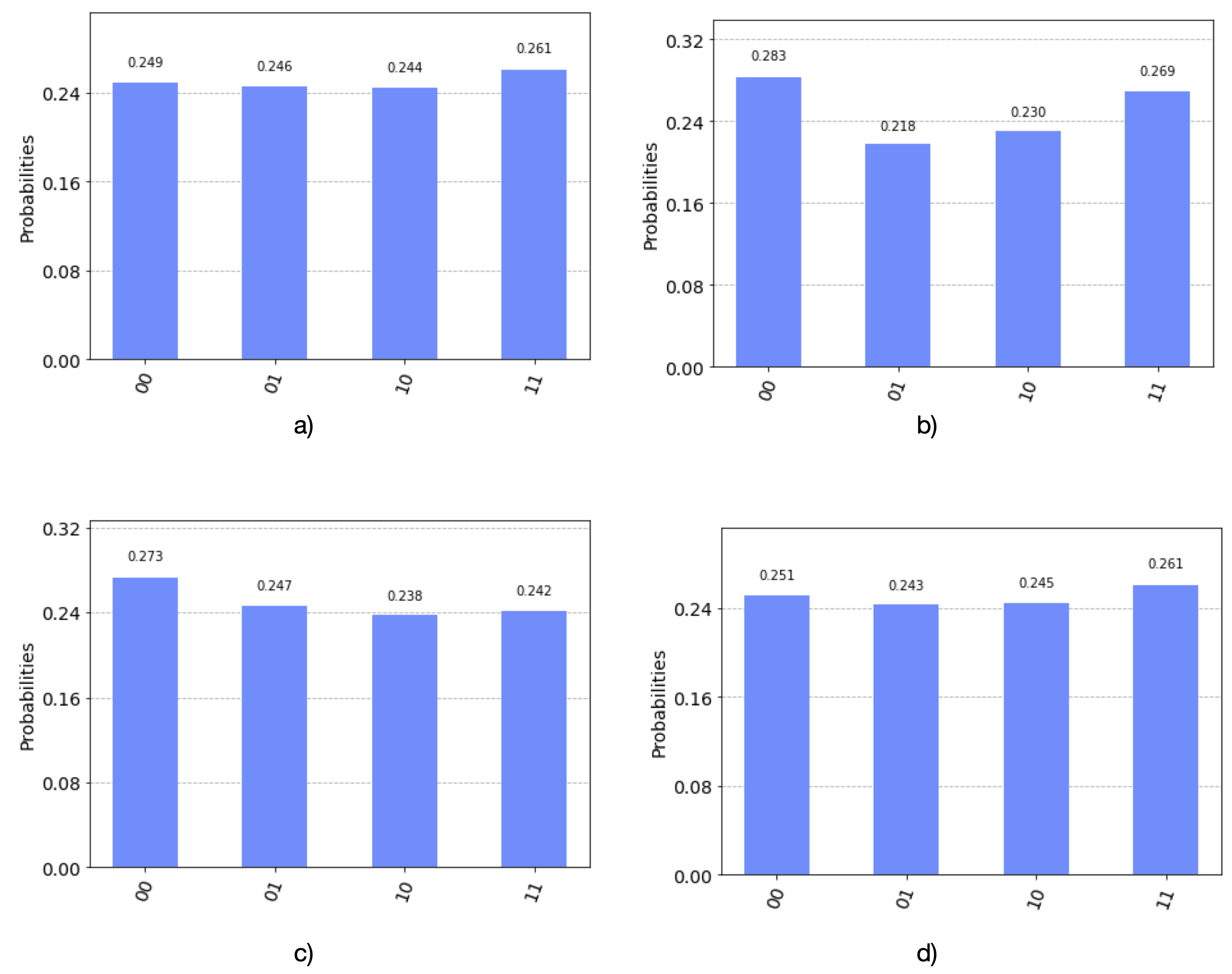}
\caption{This figure illustrates the measurement results in the form of plot histogram of the state $\hat{H}|r\rangle$ for every $r \in \{00,01,10,11\}$. Each corresponding circuit was executed 20,000 times on the IBM Manila Quantum computer. (\textbf{a}) Plot histogram of measurement results of the state $|\phi\rangle = \hat{H}|00\rangle.$ (\textbf{b}) Plot histogram of measurement results of the state $|\phi\rangle = \hat{H}|01\rangle.$ (\textbf{d}) (\textbf{c}) Plot histogram of measurement results of the state $|\phi\rangle = \hat{H}|10\rangle.$ (\textbf{d}) Plot histogram of measurement results of the state $|\phi\rangle = \hat{H}|11\rangle.$}
\label{Fig11}
\end{figure}

In Qiskit, The Operator class is used to represent matrix operators acting on a quantum system. Qiskit allows to create an operator object by initializing it with a matrix given as a list or a Numpy array. Thus we define an operator \verb|superposition_operator| by specifying the underlying matrix $\hat{H}$ row by row as \verb|superposition_operator = Operator([[1/2,1/2,-1/2,-1/2], [1/2,-1/2,|
\noindent \verb|(-1j)/2,(1j)/2], [1/2,-1/2,(1j)/2,(-1j)/2],|
\noindent\verb|[1/2,1/2,1/2,1/2]])|. To test the \verb|superposition_operator| we have created four quantum circuits in Qiskit with 2 qubits and 2 classical bits each. The state of each one of this circuits has been initialized to one of the four possible states vectors of the system, namely, $|00\rangle, |01\rangle, |10\rangle,$ and $|11\rangle$. We then applied \verb|superposition_operator| to the initial state of each of the four circuits, and measured the results. Each such circuit has been executed on both, the IBM Qasm simulator, and the IBM Manila quantum computer 20,000 times. We illustrate the measurement results in a form of plot histograms in Figure~\ref{Fig10} and Figure~\ref{Fig11} corresponding to the simulated results and results of the circuit executed on a quantum computer respectively.

In this work, we decided to use the operator $\hat{H}$ to create superposition states, since this operator was discussed in our previous work~\cite{EPJ}, and will be used in our future work to generate quantum raw random numbers. 

\subsection{Dispatching}

The dispatching procedure used in this work differs slightly from the one described in our previous work~\cite{ICCCAS,JCM}. In~\cite{ICCCAS,JCM} we simply use Python's \verb|seed| and \verb|randint| functions to randomly choose a few  Permutation Operations from the Permutation Pad and Inverse Permutation Pad by index. That is, we chose a few indices that correspond to the positions of the Permutation Operators in the Permutation Pad and its Hermitian conjugates in the Inverse Permutation Pad respectively. One of these Permutation Operators acts on a given plaintext state and its Hermitian conjugate acts on the ciphertext state respectively. 

The dispatching procedure described in this work is similar. The pre-shared key was used to create a list of indices, where each index indicates the position of the Permutation Operator in the Permutation Pad and its Hermitian conjugate in the Inverse Permutation Pad respectively. Then, based on the index, a single Permutation Operator is dispatched from the Permutation Pad and Inverse Permutation Pad to encrypt or decrypt the corresponding state. That is, one operator per state.

\subsection{Encryption with Quantum Permutation Pad}\label{sec:ADencryption}
In this work, the state vector of the system, before the encryption step with QPP, is of the form 
\begin{equation}
    |\phi\rangle = \alpha_{a}|a\rangle + \alpha_{b}|b\rangle + \alpha_{c}|c\rangle + \alpha_{d}|d\rangle, 
\end{equation} where the amplitudes satisfy $|\alpha_{a}|^2 + |\alpha_{ b}|^2 + |\alpha_{c}|^2 +|\alpha_{d}|^2 = 1$, and $a,b,c,d \in \{00, 01, 10, 11\}$.
In Section~\ref{sec:createsuperposition}, we discussed the possible superposition states that are produced by applying the operator $\hat{H}$ to the initial state vector of any given circuit. 
The produced states form the set 
\begin{align}\label{eq:setS}
    &S = \{\frac{1}{2}|00\rangle + \frac{1}{2}|01\rangle +\frac{1}{2}|10\rangle+\frac{1}{2}|11\rangle, \frac{1}{2}|00\rangle - \frac{1}{2}|01\rangle -\frac{1}{2}|10\rangle+\frac{1}{2}|11\rangle,\\ \nonumber &\frac{1}{2}|00\rangle - \frac{1}{2}i|01\rangle +\frac{1}{2}i|10\rangle+\frac{1}{2}|11\rangle,-\frac{1}{2}|00\rangle + \frac{1}{2}i|01\rangle -\frac{1}{2}i|10\rangle+\frac{1}{2}|11\rangle\}. 
\end{align}
So for any $a,b,c,d \in \{00,01,10,11\}$, the state $|\phi\rangle$ is in the set $S$. 

The Permutation Operations are linear operations as shown in~\cite{QPP}. Thus, applying them to superposition states creates new superposition states. Indeed, let $P$ be a Permutation Operator dispatched from the Permutation Pad, then 
\begin{align}
    P|\phi\rangle = &P(\alpha_{a}|a\rangle + \alpha_{b}|b\rangle + \alpha_{c}|c\rangle + \alpha_{d}|d\rangle) = \\ \nonumber
    &\alpha_{a}P|a\rangle + \alpha_{b}P|b\rangle + \alpha_{c}P|c\rangle + \alpha_{d}P|d\rangle = \\ \nonumber
    &\alpha_{a}|a'\rangle + \alpha_{b}|b'\rangle + \alpha_{c}|c'\rangle + \alpha_{d}|d'\rangle,
\end{align} where $a', b', c', d' \in \{00,01,10,11\}$. Thus, Permutation Operators essentially reassign the phases of the qubits, which leads to a few notable observations. For instance let $P$ be an operator such that $P|00\rangle = |01\rangle, P|01\rangle = |10\rangle$, $P|10\rangle = |11\rangle$ and $P|11\rangle = |00\rangle$. Then, 
\begin{align}
P(\hat{H}|01\rangle) = &P(\frac{1}{2}|00\rangle - \frac{1}{2}|01\rangle -\frac{1}{2}|10\rangle+\frac{1}{2}|11\rangle) = \\ \nonumber
&\frac{1}{2}P|00\rangle - \frac{1}{2}P|01\rangle -\frac{1}{2}P|10\rangle+\frac{1}{2}P|11\rangle =\\ \nonumber
&\frac{1}{2}|01\rangle - \frac{1}{2}|10\rangle -\frac{1}{2}|11\rangle+\frac{1}{2}|00\rangle \notin S,
\end{align} even up to a global phase, where $S$ is the set of all possible superposition states given in the Eq.~\eqref{eq:setS}. That is, $P(\hat{H}|01\rangle) \ne e^{i\theta}\hat{H}|r\rangle$ for any $r \in \{00,01,10,11\}$ and any global phase.

On the other hand, let the Permutation Operator be such that $P|00\rangle = |10\rangle, P|01\rangle = |11\rangle$, $P|10\rangle = |00\rangle$ and $P|11\rangle = |01\rangle$. Then, 
\begin{align}
P(\hat{H}|01\rangle) = &P(\frac{1}{2}|00\rangle - \frac{1}{2}|01\rangle -\frac{1}{2}|10\rangle+\frac{1}{2}|11\rangle) = 
\\ \nonumber
&-\frac{1}{2}|00\rangle+\frac{1}{2}|01\rangle + \frac{1}{2}|10\rangle - \frac{1}{2}|11\rangle \in S \text{ up to a global phase}.
\end{align}  So applying the operator $P$, previously specified, to the state $\hat{H}|01\rangle$ produced a new superposition state which is equivalent to the state $\hat{H}|01\rangle$ up to a global phase. In other words, the encryption did not essentially change the pre-encryption state.

Another notable observation is the so-called super superposition state that remains the same after the encryption, including the global phase. Given the same Permutation Operator $P$, consider 
\begin{align}
P(\hat{H}|00\rangle) = &P(\frac{1}{2}|00\rangle + \frac{1}{2}|01\rangle +\frac{1}{2}|10\rangle+\frac{1}{2}|11\rangle) = \\ \nonumber
&\frac{1}{2}|00\rangle+\frac{1}{2}|01\rangle + \frac{1}{2}|10\rangle + \frac{1}{2}|11\rangle  = \hat{H}|00\rangle \in S,
\end{align} where $S$ is the set of all possible superposition states given in the Eq.~\eqref{eq:setS}. This phenomenon demonstrates a significant difference between the encryption of basis states and superposition states using QPP. When encrypting basis states, any given single state with phase $\alpha =1$ is changed and a new state with the same phase $\alpha =1$ is produced. On the other hand, when encrypting the superposition states, the phases are reassigned and in certain cases, the produced state is equivalent to the input state. In Section~\ref{sec:Security} we discuss the security implications of this phenomenon. We point out that this phenomenon is introduced by the operator $\hat{H}$ and does not affect the QPP algorithm itself. Moreover, the adversary can not use this feature to perpetrate an attack since the adversary does not have knowledge of the permutation operator used. In other words, the adversary does not know whether a given ciphertext is equal to the pre-encryption state or not. 

\begin{figure}[ht]
\centering
\includegraphics[width=12 cm]{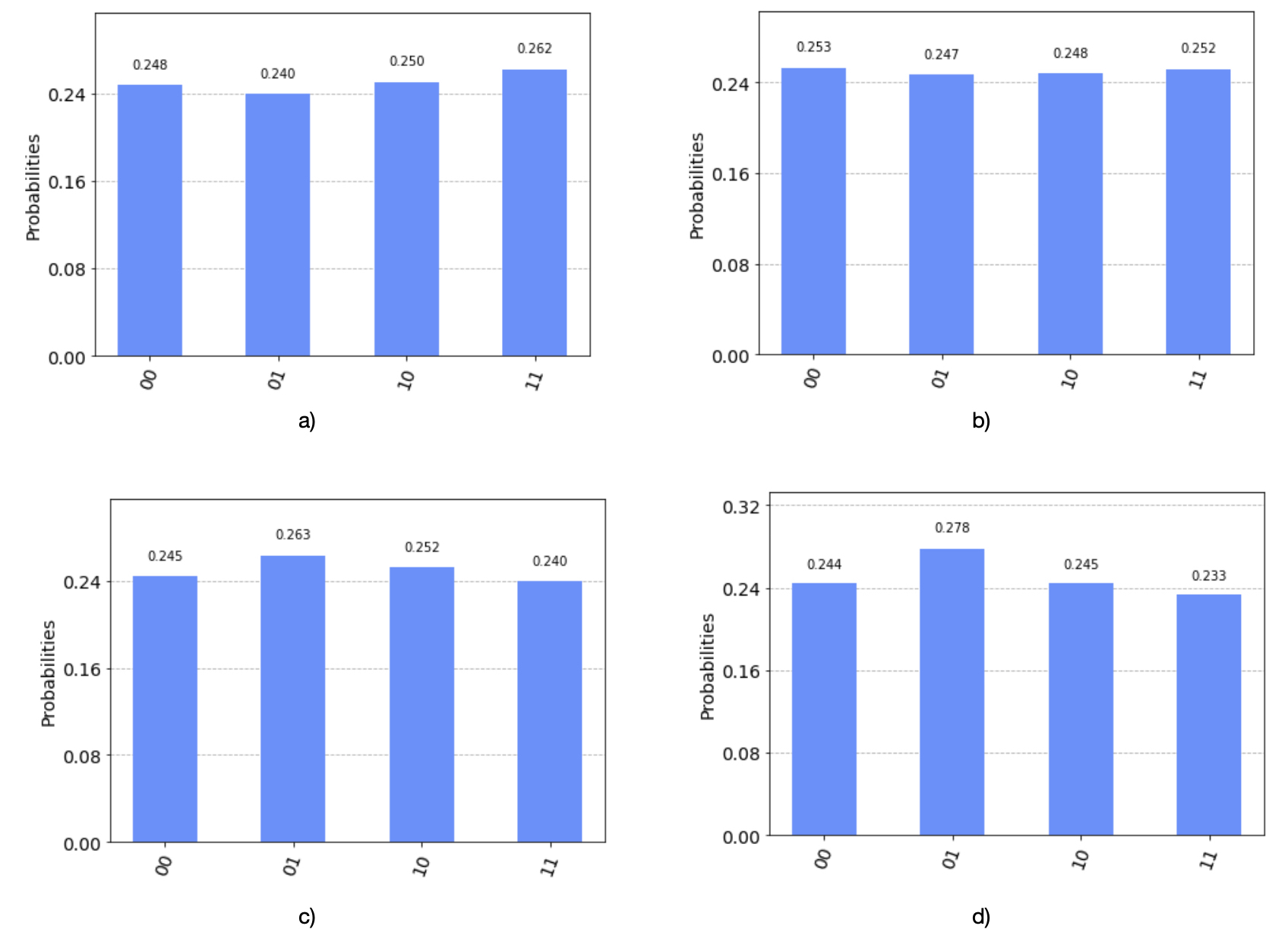}
\caption{This figure illustrates the measurement results in the form of plot histogram to verify that any given ciphertext state is a superposition of four states. The results are illustrated for some randomly chosen Permutation Operator $P$. The circuit was executed on the IBM Qasm simulator 20,000 times. (\textbf{a}) Plot histogram of measurement results of the state $|\phi\rangle = P(\hat{H}|00\rangle).$ (\textbf{b}) Plot histogram of measurement results of the state $|\phi\rangle = P(\hat{H}|01\rangle).$ (\textbf{c}) Plot histogram of measurement results of the state $|\phi\rangle = P(\hat{H}|10\rangle).$ (\textbf{d}) Plot histogram of measurement results of the state $|\phi\rangle = P(\hat{H}|11\rangle).$}
\label{Fig12}
\end{figure}

\begin{figure}[ht]
\centering
\includegraphics[width=12 cm]{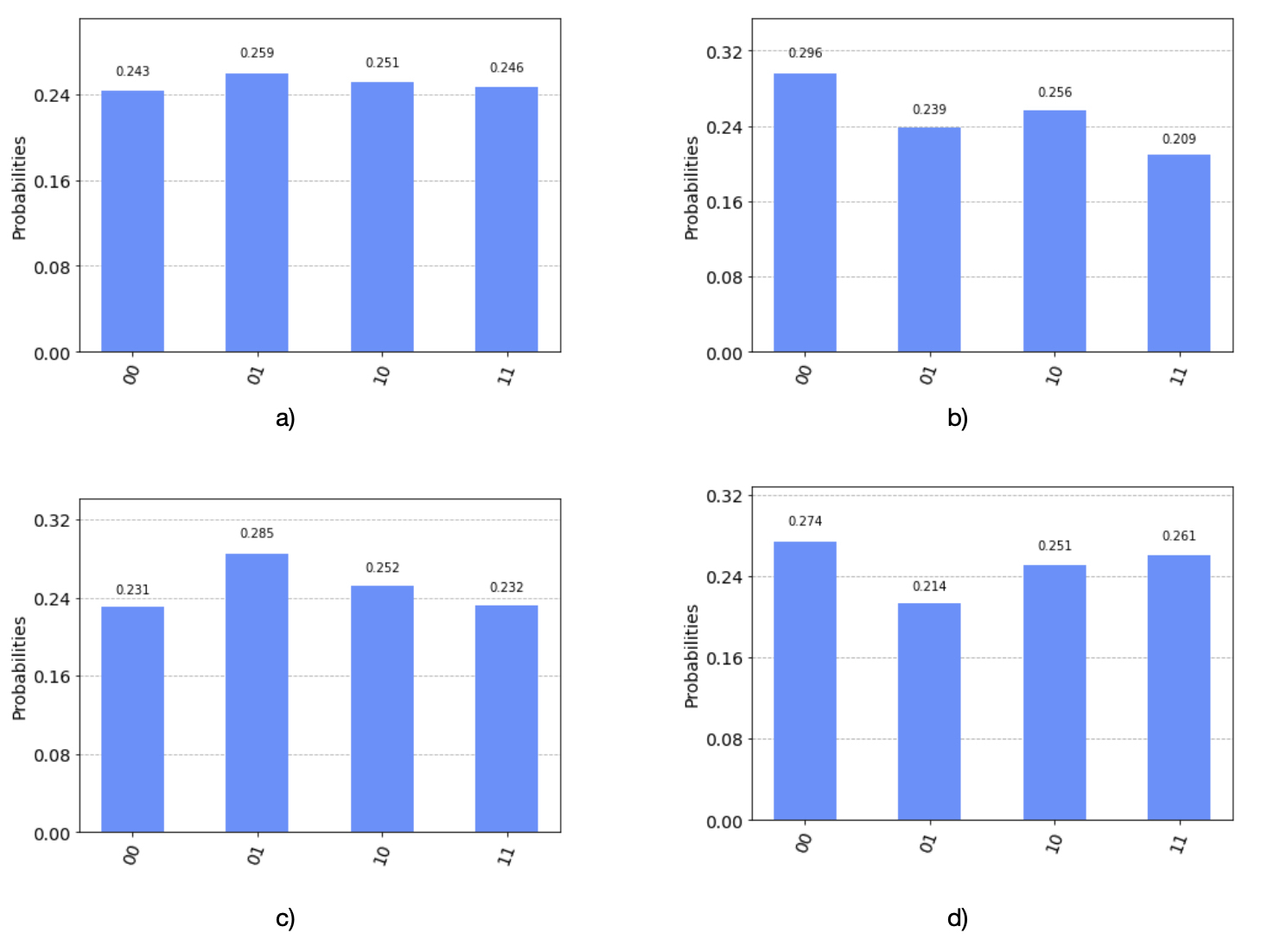}
\caption{This figure illustrates the measurement results in the form of plot histogram to verify that any given ciphertext state is a superposition of four states. The results are illustrated for some randomly chosen Permutation Operator $P$. The circuit was executed on the IBM Manila Quantum computer 20,000 times. (\textbf{a}) Plot histogram of measurement results of the state $|\phi\rangle = P(\hat{H}|00\rangle).$ (\textbf{b}) Plot histogram of measurement results of the state $|\phi\rangle = P(\hat{H}|01\rangle).$ (\textbf{c}) Plot histogram of measurement results of the state $|\phi\rangle = P(\hat{H}|10\rangle).$ (\textbf{d}) Plot histogram of measurement results of the state $|\phi\rangle = P(\hat{H}|11\rangle).$}
\label{Fig13}
\end{figure}

We provide the reader with Figure~\ref{Fig12} and Figure~\ref{Fig13} depicting the measurement results of the ciphertext states $P(\hat{H}|r\rangle)$, for every $r \in \{00,01,10,11\}$ and a randomly chosen Permutation Operator $P$ to demonstrate that the ciphertext states are a superposition of four states. Figure~\ref{Fig12} corresponds to a sample circuit executed on the IBM Qasm simulator 20,000 times, and Figure~\ref{Fig13} corresponds to the same circuit executed on the IBM Manila Quantum computer 20,000 times.


\subsection{Decryption of superposition states}\label{sec:decrypt}
Following the general step-by-step logic of how QPP is used to encrypt superposition states, illustrated in Figure~\ref{Fig5}, it is clear that decryption can be done by applying Hermitian conjugates of the operators used for encryption in the correct order. Indeed, the operator that creates superposition states is a unitary. It was proved in~\cite{QPP} that Permutation Operators are unitary operators. The XOR operation used for randomization can be used again with the same key for de-randomization. Thus, it is possible to apply Permutation Operators from the Inverse Permutation Pad, followed by the Hermitian conjugate of the operator $\hat{H}$, followed by the XOR operator with the same key to decrypt the ciphertext. Indeed, let $P$ denote a Permutation Operator, then for each randomized $2$-bit plaintext block $r$ we have 
\begin{equation}
 \hat{H}^{\dagger}(P^{\dagger}(P(\hat{H}(|r\rangle)))) = \hat{H}^{\dagger}(P^{\dagger}P)\hat{H}|r\rangle = \hat{H}^{\dagger}\hat{H}|r\rangle = |r\rangle.
\end{equation} Suppose that $r = m \oplus k$, where $m$ denotes the message, and $k$ denotes the key. Then, we have 
\begin{equation}
    r \oplus k = (m \oplus k) \oplus k = m \oplus (k \oplus k) = m.
\end{equation}

\begin{figure}[ht]
\centering
\includegraphics[width=16cm]{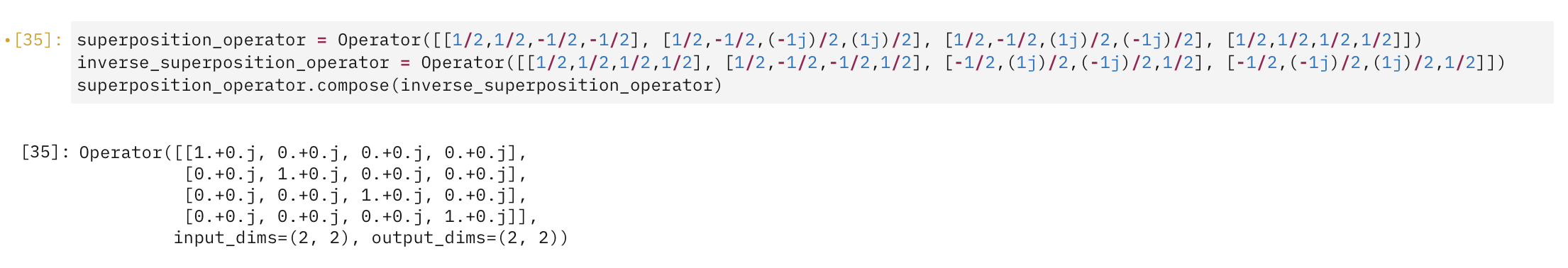}
\caption{An illustration of a Qiskit command that verifies that the operator $\hat{H}^{\dagger}$ is indeed the Hermitian conjugate of the operator $\hat{H}$.}
\label{Fig14}
\end{figure}

There are different ways to create Hermitian conjugate operators in Qiskit. One way that we used in our previous work~\cite{ICCCAS} is to create inverse circuit consisting of the operator, for which we are looking to create Hermitian conjugate. Such circuit can be appended to the main circuit in the appropriate place. This way, although is correct, is not the most convenient. Far better way is to create conjugate transpose of the matrices and use them to create Qiskit Operator objects. In Section~\ref{sec:createsuperposition} we discuss how to create an operator by specifying the corresponding matrices row-by-row. We used this technique to create Hermitian conjugate operators. The source code for this part can be found in the Appendix.  We give an example of the 

\begin{lstlisting}[language = Python]
inverse_superposition_operator = Operator([[1/2,1/2,1/2,1/2], [1/2,-1/2,-1/2,1/2], [-1/2,(1j)/2,(-1j)/2,1/2], [-1/2,(-1j)/2,(1j)/2,1/2]])
\end{lstlisting}

where \verb|inverse_superposition_operator| denotes the operator $\hat{H}^{\dagger}.$
The reader is welcome to test that this operator is indeed the Hermitian conjugate of the \verb|superposition_| \verb|operator|, using for instance, the procedure illustrated in Figure~\ref{Fig14}.

Overall, to guarantee the successful decryption we rely on the property $U^{\dagger}U = I$, for a unitary operator $U$, and the properties of the classical XOR function. In the ideal case, where randomization is done using the CX gates, the same property $U^{\dagger}U = I$ can be used for decryption since CX is a unitary operator. 

\begin{figure}[ht]
\centering
\includegraphics[width=12 cm]{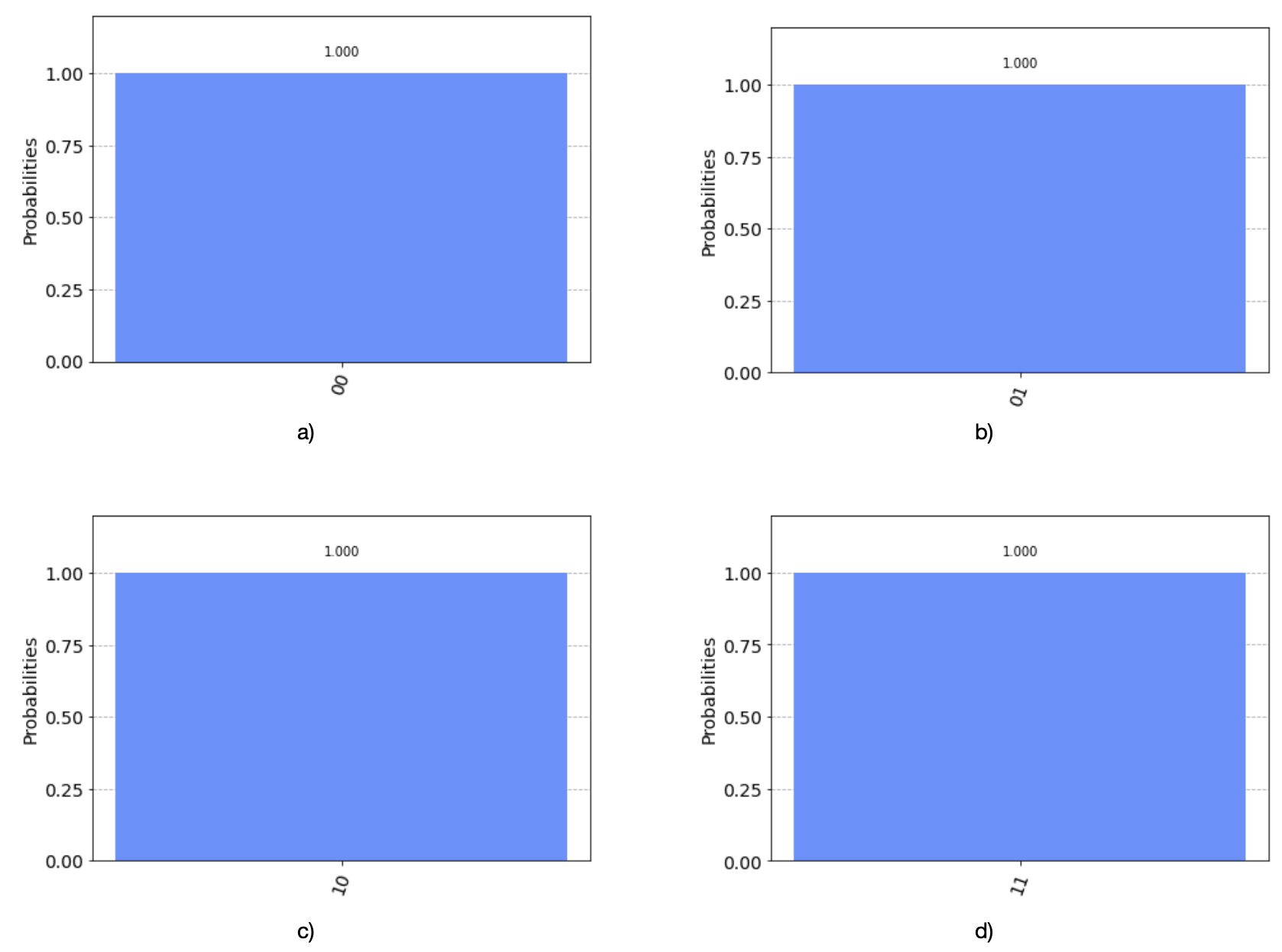}
\caption{This figure illustrates measurement results of the state $\hat{H}^{\dagger}(P^{\dagger}(P(\hat{H}|r\rangle))),$ for some randomly chosen Permutation Operator $P$ and $r \in \{00,01,10,11\}.$ Each circuit was executed 20,000 times on the IBM Qasm simulator. This figure illustrates the measurement results in the form of plot histogram. (\textbf{a}) Plot histogram of measurement results of the state $|\phi\rangle = \hat{H}^{\dagger}(P^{\dagger}(P(\hat{H}|00\rangle))).$ (\textbf{b}) Plot histogram of measurement results of the state $|\phi\rangle = \hat{H}^{\dagger}(P^{\dagger}(P(\hat{H}|01\rangle))).$ (\textbf{c}) Plot histogram of measurement results of the state $|\phi\rangle = \hat{H}^{\dagger}(P^{\dagger}(P(\hat{H}|10\rangle))).$ (\textbf{d}) Plot histogram of measurement results of the state $|\phi\rangle = \hat{H}^{\dagger}(P^{\dagger}(P(\hat{H}|11\rangle))).$}
\label{Fig15}
\end{figure}

\begin{figure}[ht]
\centering
\includegraphics[width=12 cm]{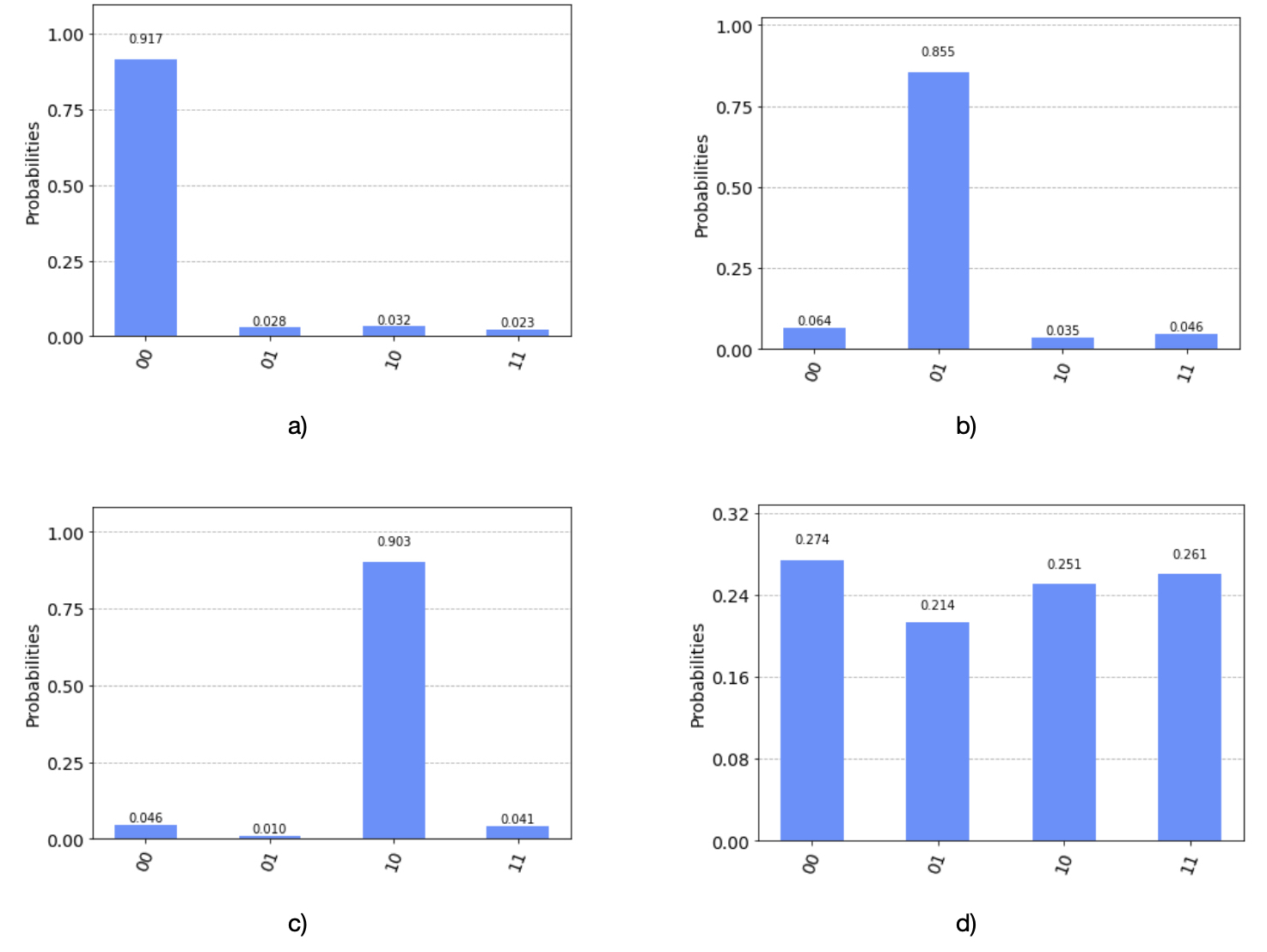}
\caption{This figure illustrates measurement results of the state $\hat{H}^{\dagger}(P^{\dagger}(P(\hat{H}|r\rangle))),$ for some randomly chosen Permutation Operator $P$ and $r \in \{00,01,10,11\}.$ Each circuit was executed 20,000 times on the IBM Manila Quantum computer. This figure illustrates the measurement results in the form of plot histogram. (\textbf{a}) Plot histogram of measurement results of the state $|\phi\rangle = \hat{H}^{\dagger}(P^{\dagger}(P(\hat{H}|00\rangle))).$ (\textbf{b}) Plot histogram of measurement results of the state $|\phi\rangle = \hat{H}^{\dagger}(P^{\dagger}(P(\hat{H}|01\rangle))).$ (\textbf{c}) Plot histogram of measurement results of the state $|\phi\rangle = \hat{H}^{\dagger}(P^{\dagger}(P(\hat{H}|10\rangle))).$ (\textbf{d}) Plot histogram of measurement results of the state $|\phi\rangle = \hat{H}^{\dagger}(P^{\dagger}(P(\hat{H}|11\rangle))).$}
\label{Fig16}
\end{figure}

To demonstrate that the decryption procedure works indeed, as specified in this section we provide the reader with Figure~\ref{Fig15} and Figure~\ref{Fig16} illustrating the measurement result of the state $\hat{H}^{\dagger}(P^{\dagger}(P(\hat{H}|r\rangle)))$, for a randomized plaintext block $r \in \{00,01,10,11\}.$

\subsection{Security}\label{sec:Security}
The security of the QPP algorithm is discussed in detail in~\cite{QPP,qpp-kuang2020}. In fact, Kuang and Bettenburg showed that the QPP algorithm as described in~\cite{QPP} achieves perfect secrecy~\cite{qpp-kuang2020}. In this work, we apply the encryption and decryption procedures on the superposition states. Indeed, the superposition states are created and respectively transformed back into the basis states before the encryption and respectively after the decryption, without changing the QPP encryption and decryption mechanisms. 
 We do not aim to discuss the in-depth security of the current implementation, as it deserves a separate paper, however, we will briefly mention certain security aspects that we found interesting. 

Note that there are advantages and disadvantages of having the ciphertext states being superposition states. Since the only way for the attacker to obtain the ciphertext is to measure it and since the measurement destroys the superposition states, the attacker, in general, will not be able to obtain the ciphertext in its original superposition form, but rather just the corresponding measurement outcome. Meanwhile, when the ciphertext is a basis state, the adversary can learn the ciphertext directly. Thus, having the ciphertext being a superposition state disfavors the attacker and adds an extra layer of security to the scheme. 

On the other hand, as discussed in Section~\ref{sec:createsuperposition}, encryption of the superposition states with Permutation Operators reassigns the phases of the qubits. Thus, we must consider the case that the adversary can act on the ciphertext by acting on the phases alone and try to reassign them back. This will require the adversary to act on the ciphertext in its original form without measurement, which is not generally possible. 

 Recall that, unlike the Permutation Operators that are chosen to be applied at random, the single operator $\hat{H}$ is applied to each one of the initial state vectors of every encryption circuit. Moreover, the operator $\hat{H}$ is publicly shared. Thus, it is natural to examine whether knowing $\hat{H}$ leaks any information about the plaintext or the secret key. We claim that knowing the operator $\hat{H}$ does not benefit the attacker. 

 Suppose that the ciphertext is in the state $$P(\hat{H}|00\rangle) = \frac{1}{2}|00\rangle + \frac{1}{2}|01\rangle +\frac{1}{2}|10\rangle+\frac{1}{2}|11\rangle,$$ which is indistinguishable from the state $$P(\hat{H}|01\rangle) = \frac{1}{2}|01\rangle - \frac{1}{2}|10\rangle -\frac{1}{2}|11\rangle+\frac{1}{2}|00\rangle$$ from the point of view of the attacker. Since the attacker can not observe the superposition state, but only measure it, the adversary can not distinguish between the two. Indeed, denote the measurement result as $m_0$. It is true that $$Pr(m_0||\phi\rangle = P(\hat{H}|00\rangle)) = Pr(m_0||\phi\rangle = P(\hat{H}|01\rangle)) = \frac{1}{4}.$$
 In the first case, applying the operator $\hat{H}$ to the ciphertext will yield the randomized plaintext block $``00"$. Indeed, for any Permutation Operator $P$
 \begin{align}
P(\hat{H}|00\rangle) = &P(\frac{1}{2}|00\rangle + \frac{1}{2}|01\rangle +\frac{1}{2}|10\rangle+\frac{1}{2}|11\rangle) = \\ \nonumber
&\frac{1}{2}|a\rangle + \frac{1}{2}|b\rangle +\frac{1}{2}|c\rangle+\frac{1}{2}|d\rangle = \hat{H}|00\rangle,
\end{align} 
where $a,b,c,d \in \{00,01,10,11\}.$ Thus, 
\begin{equation}
    \hat{H}^{\dagger}[P(\hat{H}|00\rangle )] = |00\rangle,
\end{equation} for any Permutation Operator $P$. Meanwhile, applying $\hat{H}$ to the ciphertext $P(\hat{H}|01\rangle) = \frac{1}{2}|01\rangle - \frac{1}{2}|10\rangle -\frac{1}{2}|11\rangle+\frac{1}{2}|00\rangle$ will further obscure the plaintext. Consider
$$\hat{H}^{\dagger}[P(\hat{H}|01\rangle)] = \frac{1}{4}\begin{bmatrix}
0\\
0\\
-2+2i\\
-2-2i
\end{bmatrix}.$$
Measuring both states will not yield any information about the plaintext, moreover, the adversary won't be able to distinguish the original form of the ciphertext or tell which permutation operator $P$ was used for encryption.

Another point that we would like to raise is that, in general, having ciphertext as a superposition state disables the attacker to act on it in its original form. Indeed, if the ciphertext state is a superposition state, the attacker needs to measure it which will collapse the state to a basis state.

\begin{figure}[ht]
\centering
\includegraphics[width=12 cm]{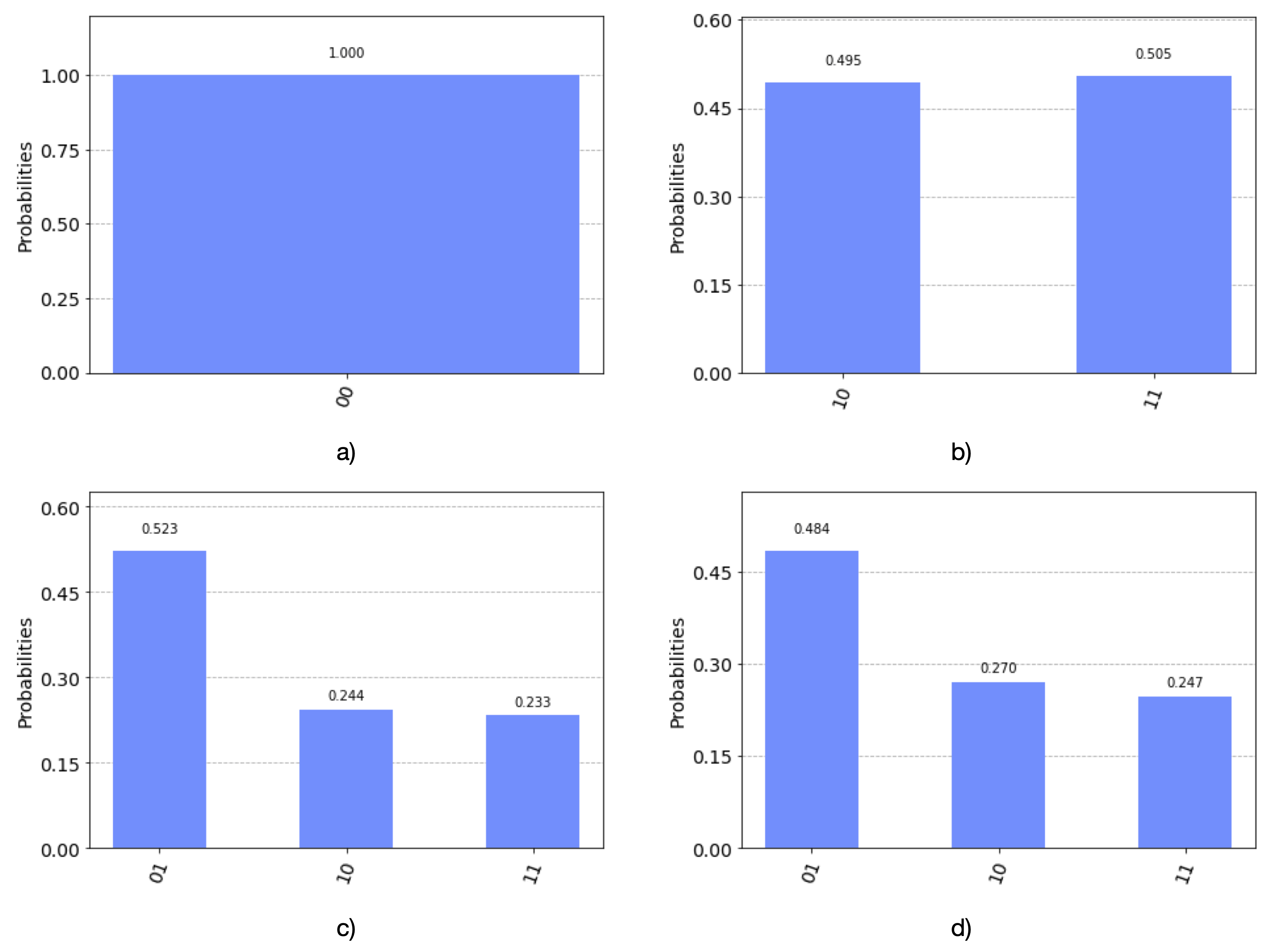}
\caption{This figure illustrates the measurement results in the form of plot histogram. Each corresponding circuit was executed 20,000 times on the IBM Qasm simulator. (\textbf{a}) Plot histogram of measurement results of the state $|\phi\rangle = \hat{H}^{\dagger}(P(\hat{H}|00\rangle)).$ (\textbf{b}) Plot histogram of measurement results of the state $|\phi\rangle = \hat{H}^{\dagger}(P(\hat{H}|01\rangle)).$ (\textbf{c}) Plot histogram of measurement results of the state $|\phi\rangle = \hat{H}^{\dagger}(P(\hat{H}|10\rangle)).$ (\textbf{d}) Plot histogram of measurement results of the state $|\phi\rangle = \hat{H}^{\dagger}(P(\hat{H}|11\rangle)).$}
\label{Fig17}
\end{figure}

\begin{figure}[ht]
\centering
\includegraphics[width=12 cm]{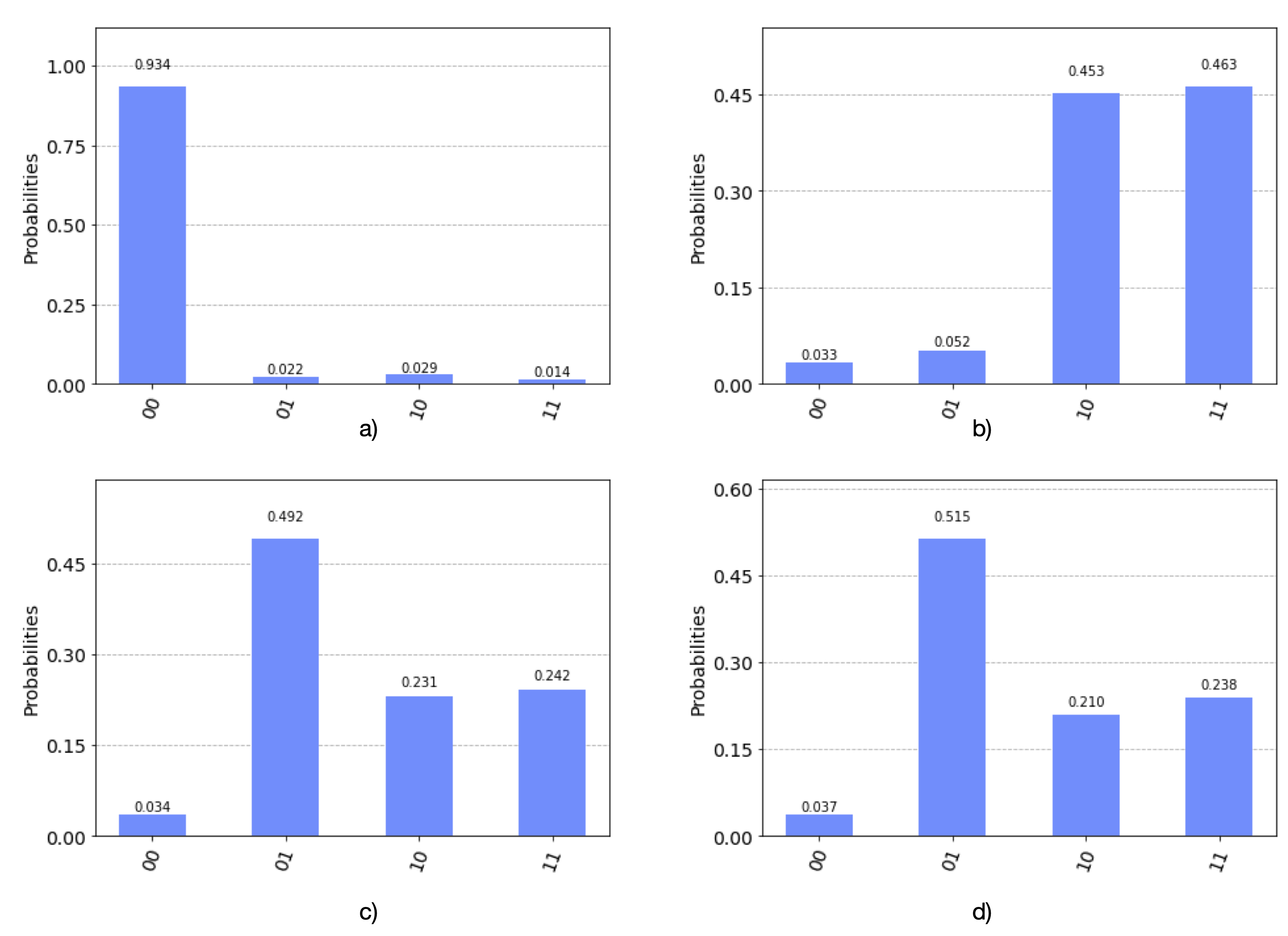}
\caption{ This figure illustrates the measurement results in the form of plot histogram. Each corresponding circuit was executed 20,000 times on the IBM Manila Quantum computer. (\textbf{a}) Plot histogram of measurement results of the state $|\phi\rangle = \hat{H}^{\dagger}(P(\hat{H}|00\rangle)).$ (\textbf{b}) Plot histogram of measurement results of the state $|\phi\rangle = \hat{H}^{\dagger}(P(\hat{H}|01\rangle)).$ (\textbf{c}) Plot histogram of measurement results of the state $|\phi\rangle = \hat{H}^{\dagger}(P(\hat{H}|10\rangle)).$ (\textbf{d}) Plot histogram of measurement results of the state $|\phi\rangle = \hat{H}^{\dagger}(P(\hat{H}|11\rangle)).$}
\label{Fig18}
\end{figure}

We illustrate measurement results of the sample states $\hat{H}^{\dagger}(P(\hat{H}|r\rangle))$, for a randomly chosen Permutation Operator $P$ and every $r \in \{00,01,10,11\}$ in Figure~\ref{Fig17} and Figure~\ref{Fig18}.

\subsection{QPP and other Quantum Encryption Schemes}
A reader familiar with Quantum encryption will notice similarities between the QPP algorithm and the Quantum One-Time Pad algorithm (QOTP)~\cite{QOTP,MoscaQOTP,Leung}. Boykin and Roychowdhury pointed out that a generic quantum algorithm to encrypt quantum data would consist of a finite number of unitary operators $U_k$, used to encrypt a quantum state and their respective Hermitian conjugates $U_k^{\dagger}$ are used for decryption, utilizing the property $U_k^{\dagger}U_k|r\rangle = |r\rangle$~\cite{QOTP}. More explicitly, the value $k$ is a classical secret key. The key $k$ specifies the unitary operator $U_k$ that is applied to encrypt a quantum state that represents a message. To decrypt the produced ciphertext, an operator $U_k^{\dagger}$ is applied to the ciphertext state to retrieve the original state~\cite{QOTP}. Inspired by the property $U_k^{\dagger}U_k|r\rangle = |r\rangle$, QPP was developed with specific unitary operators in mind, namely, Permutation operators. Note that while both QPP and QOTP schemes use classical key material to determine whether to act on the plaintext qubits, and what unitary operators are acting on the given quantum state, the techniques used in the QOTP scheme are different. For instance, we have not necessarily considered permutation operators from an angle of basis gates. We use the Qiskit \verb|transpile()| to express permutation operators in terms of basis gates with no particular conditions, while the QOTP scheme uses basis operators $\sigma_{x}$ and $\sigma_{z}$. Moreover, QPP is designed to act on basis states, producing basis states as ciphertext, while in the framework of the QOTP scheme, the ciphertext is a totally mixed quantum state. We also mention that QOTP requires $2n$ classical bits for encryption, while the size of the key material and the number of qubits in the framework of QPP depends on the entropy required. Note also that QPP can be implemented on both classical and quantum computers and is not limited by the quantum nature of the encryption operators used, while QOTP is truly a quantum encryption algorithm that is designed for quantum devices alone. 

Another quantum encryption scheme closely related to QOTP is the $\mathcal{EHE}$ scheme by Liang and Yang~\cite{EHE}. In fact, the authors refer to the QOTP as a special case of the $\mathcal{EHE}$ scheme. Note that in~\cite{EHE} quantum block encryption (QBE) scheme is constructed in the form of $\mathcal{EHE}$ encryption using two pseudorandom functions. In our work, although we break the plaintext and the secret key into blocks, QPP is not a block cipher. In this implementation, plaintext and secret key are broken into blocks due to the limitations of the current quantum computers. As the NISQ era devices improve, we will be able to implement QPP to encrypt an entire message as a whole. The $\mathcal{EHE}$ encryption is a three-part process. First, the plaintext $p$ is encrypted using the first quantum encryption scheme $\mathcal{E}_{k_1}$, producing $c_1 = \mathcal{E}_{k_1}(p)$. Next, a transversal Hadamard transformation is then performed on $c_1$, producing $c_1' = Hc_1$. The new ciphertext $c_1'$ is then encrypted again using the second quantum encryption scheme $\mathcal{E}_{k_2}$, producing $c_2 = \mathcal{E}_{k_2}H\mathcal{E}_{k_1}(p)$. The implementation of QPP described in this work follows a similar process. Indeed, the message is first encrypted using the XOR function with the secret key. The produced state is then used to create a superposition using the operator $\hat{H}$, which is then encrypted using QPP Permutation operators. However, we would like to point out that the QPP algorithm itself is not limited to superposition states. That is, this work explored whether QPP would benefit from introducing superposition states. The core of the QPP algorithm remains the encryption of a randomized plaintext with permutation operators, while in the framework of the $\mathcal{EHE}$ scheme, the use of the Hadamard gate $H$ is a significant part of the algorithm design. Liang and Yang in~\cite{EHE} considered the security of the scheme in terms of IND-CPA and IND-CCA properties as defined in~\cite{Xiang2015,Alagic_2016,Liang2020,Broadbent2015}. Note that we have not considered the current implementation in terms of IND-CPA or IND-CCA, as the focus of this work was on the implementation of the scheme. Moreover, we claim that the security of the current implementation depends largely on the choice of the operator $\hat{H}$ for the creation of the superposition states. 
\section{Conclusion}\label{sec:conclusion}

In this work, we report on an implementation of Kuang and Bettenburg's symmetric Quantum Permutation Pad (QPP) algorithm used to encrypt superposition states in IBM quantum systems using the Qiskit development kit. This work builds on and extends our previous research~\cite{ICCCAS,JCM,EPJ}, focusing on the implementation of the QPP algorithm used to encrypt basis states. The implementation described in this paper is fully-functioning, lightweight, and can be run on any IBM quantum system with at least $5$ qubits and a Quantum Volume of $32$. This implementation includes the addition of a plaintext randomization procedure which is important to the overall security of the algorithm. The focus of this work, the creation, and encryption of superposition states, was discussed in depth. We explained how these procedures are implemented and introduced an operator $\hat{H}$ established from the diagonalization of a certain permutation operator $P_1$ given in this paper. This operator $\hat{H}$ was used to create the superposition states, which were then encrypted using QPP. We also briefly discuss the security of this implementation. 

This work broadens the applicability of QPP for the encryption of both basis states and superposition states. In the future, we will examine if the implementation of QPP can be extended to entangled states. Moreover, we will continue analyzing the general security of the QPP algorithm used to encrypt superposition states.

\section{Figures}
This section provides illustrations to the procedures described above to aid in reader's understanding of the fundamentals of the QPP symmetric scheme.  

\clearpage






\section*{Appendix}
We include a source code for the implementation described in this work. Current version of the source code as well as the more detailed implementation code is available upon request to the corresponding author. 

\begin{lstlisting}[language = Python]
import numpy as np

# Importing standard Qiskit libraries
from qiskit import QuantumCircuit, transpile, Aer, IBMQ
from qiskit.tools.jupyter import *
from qiskit.visualization import *
from ibm_quantum_widgets import *
from qiskit.providers.aer import QasmSimulator

# Loading your IBM Quantum account(s)
provider = IBMQ.load_account()

#importing necessary Qiskit libraries
from qiskit import execute
from qiskit import QuantumCircuit, assemble, Aer
from qiskit.visualization import plot_histogram, plot_bloch_vector
from qiskit.extensions import RXGate, XGate, CXGate, SwapGate
from qiskit import execute
from qiskit import QuantumCircuit, ClassicalRegister, QuantumRegister
from qiskit import BasicAer
from qiskit.compiler import transpile
from qiskit.quantum_info.operators import Operator, Pauli
from qiskit.quantum_info import process_fidelity
from qiskit.quantum_info import Statevector
from random import seed
from random import randint
from qiskit import QuantumCircuit, transpile
from qiskit.providers.aer import AerSimulator
from qiskit import IBMQ
from qiskit.compiler import transpile
from qiskit.tools.monitor import job_monitor


#defining the parameters 
n = 4                #possible number of quantum states generated by 2 qubits 
num_of_bits = 448    #secret key should be 448 bits long 
bits_in_block = 8    #we break the secret key into blocks of 8 bits for Permutation/ Inverse Permutation Pad generation
num_of_qubits = 2    #we operate on 2 qubits i.e. 2 qubits is the circuit input (width of the circuit)
num_of_perm_in_pad = 56    #Permutation Pad should consist of 56 permutation to guaranetee 256 bits of entropy
pad_selection_key_size = 6 #sceret key is broken into blocks of 6 bits for dispatching procedure

#the pre-shared secret key is broken into blocks 
secret_key = '111000111100011010010000010101110...0010101011011001'
secret_key_blocks = [secret_key[i:i+bits_in_block] for i in range(0, len(secret_key), bits_in_block)]

# Fisher Yates shuffling 
key_chunks = []

def randomize (arr, n):
    # Start from the last element and swap one by one. We don't
    # need to run for the first element that's why i > 0
    for i in range(n-1,0,-1):
        # Pick a random index from 0 to i
        #j = randint(0,i+1)
        j = key_chunks[i]
 
        # Swap arr[i] with the element at random index
        arr[i],arr[j] = arr[j],arr[i]
    return arr


#Permutation Pad is created 
Permutation_Pad = []

for num_of_perm in range(num_of_perm_in_pad):
    
    my_array = []
    for num in range(n):
        my_array.append(num)
    array_of_n_num = my_array.copy()
   

    
    key_block = secret_key_blocks[num_of_perm]
    key_chunks = [key_block[i:i+num_of_qubits] for i in range(0, len(key_block), num_of_qubits)]
    for num in range(len(key_chunks)):
        key_chunks[num] = int(key_chunks[num],2)
    len_of_array = len(my_array)
    shuffled_array = randomize(my_array, len_of_array)
    
    
    matrix_of_zeros = np.zeros((n, n), dtype=int)
    my_matrix = matrix_of_zeros


    for num in range(n):
        my_matrix[array_of_n_num[num]][shuffled_array[num]] = 1
        
    Permutation_Pad.append(Operator(my_matrix))


#The Inverse Permutation Pad is created
Inverse_Permutation_Pad = []

for num_of_perm in range(num_of_perm_in_pad):
    
    my_array = []
    for num in range(n):
        my_array.append(num)
    array_of_n_num = my_array.copy()
   

    
    key_block = secret_key_blocks[num_of_perm]
    key_chunks = [key_block[i:i+num_of_qubits] for i in range(0, len(key_block), num_of_qubits)]
    for num in range(len(key_chunks)):
        key_chunks[num] = int(key_chunks[num],2)
    len_of_array = len(my_array)
    shuffled_array = randomize(my_array, len_of_array)
    
    
    matrix_of_zeros = np.zeros((n, n), dtype=int)
    my_matrix = matrix_of_zeros


    for num in range(n):
        my_matrix[array_of_n_num[num]][shuffled_array[num]] = 1
        
    Inverse_Permutation_Pad.append(Operator(my_matrix.transpose()))

#The list of indices for the dispatching procedure is created 

pad_selection_blocks = [secret_key[i:i+pad_selection_key_size] for i in range(0, len(secret_key), pad_selection_key_size)]
for num in range(len(pad_selection_blocks)):
        pad_selection_blocks[num] = (int(pad_selection_blocks[num],2)%56)

#print(pad_selection_blocks)
#print(len(pad_selection_blocks))


from PIL import Image
from io import BytesIO


#code to convert PNG "cat.png" into a bitstring
out = BytesIO()
with Image.open("newcat.png") as img:
    img.save(out, format="png")
image_in_bytes = out.getvalue()
message = "".join([format(n, '08b') for n in image_in_bytes])  

#breaking the secret key into blocks used for randomization
key_for_xor_list = []
for x in range(len(message)):
    key_for_xor_list.append(secret_key[x%len(secret_key)])
key_for_xor = "".join(key_for_xor_list)

#creating the randomized plaintext 
randomized_message=[str(int(message[i])^int(key_for_xor[i])) for i in range(len(message))]
randomized_message=''.join(randomized_message)

chunk_size = num_of_qubits           #the size of each block is 2 bits/qubits
#breaking the message into blocks of 2             
message_chunks = [randomized_message[i:i+chunk_size] for i in range(0, len(randomized_message), chunk_size)] 


#creating the superposition operator
superposition_operator = Operator([[1/2,1/2,-1/2,-1/2], [1/2,-1/2,(-1j)/2,(1j)/2], [1/2,-1/2,(1j)/2,(-1j)/2], [1/2,1/2,1/2,1/2]])
inverse_superposition_operator = Operator([[1/2,1/2,1/2,1/2], [1/2,-1/2,-1/2,1/2], [-1/2,(1j)/2,(-1j)/2,1/2], [-1/2,(-1j)/2,(1j)/2,1/2]])

#Encryption procedure
my_list = []

for x in range(len(message_chunks)):
    state_vector = message_chunks[x]
    qc = QuantumCircuit(num_of_qubits, num_of_qubits)     #establish a circuit with 2 qubits and 2 classical bits to store measureemnt results later 
# read qubits from right to left so '110' will be q[0] = 0, q[1] = 1, q[2]=1
    qc.initialize(Statevector.from_label(state_vector))
    qc.barrier()
    qc.append(superposition_operator, range(num_of_qubits))
    qc.barrier
# apply gates bottom to top 


# appy the permutations two qubits at a time from top register to the bottom register using append command 
#first permutation in the list applies to top two qubits 
    j = pad_selection_blocks[x%len(pad_selection_blocks)]    #or you can apply permutations chosen from the pad using key blocks of 6 bits
    qc.append(Permutation_Pad[j], range(num_of_qubits))      #apply permutation 
    qc.barrier()
    
#here we pretend they got transmitted 
#Decryption procedure
    
    qc.append(Inverse_Permutation_Pad[j],range(num_of_qubits))
    qc.barrier()
    
    qc.append(inverse_superposition_operator, range(num_of_qubits))
    
    
    
    qc.measure([0,1], [0,1])     #measure the result 
    

#execute the circuit and store the final state in the list   
# on a simulator 
    
    simulator = Aer.get_backend('qasm_simulator')
    result = execute(qc, backend = simulator, shots = 1024).result()
    counts = result.get_counts()
    my_list.append(counts.most_frequent())   #will only append results with highest probability 

#or alternatively on a quantum computer 

    #provider = IBMQ.load_account()
    #qcomp = provider.get_backend('ibmq_manila')
    #job = execute(qc, backend=qcomp, shots = 20000)
    #job_monitor(job)
    #result = job.result()
    #counts = result.get_counts()
    #my_list.append(counts.most_frequent())


#de-randomize the obtained decrypted ciphertext

randomized_decrypted_cipher = "".join(my_list)
decrypted_message=[str(int(randomized_decrypted_cipher[i])^int(key_for_xor[i])) for i in range(len(randomized_decrypted_cipher))]
decrypted_message=''.join(decrypted_message)   

#create a png file from the obtained plaintext 

decrypted_bytes = [int(decrypted_message[i:i + 8], 2) for i in range(0, len(decrypted_message), 8)]

with open('decrypted_cat.png', 'wb') as f:
    f.write(bytes(decrypted_bytes))
   
\end{lstlisting}

\end{document}